\pdfoutput=1
\documentclass[11pt]{article}

\usepackage[final]{acl}

\usepackage{times}
\usepackage{latexsym}

\usepackage[T1]{fontenc}
\usepackage[utf8]{inputenc}

\usepackage{microtype}

\usepackage{inconsolata}

\usepackage{graphicx}

\usepackage{algorithm}
\usepackage{algorithmicx}
\usepackage{makecell}
\usepackage[most]{tcolorbox}
\usepackage[utf8]{inputenc}
\usepackage{booktabs}
\usepackage{subfig}
\usepackage{amsmath}
\usepackage{enumitem}
\usepackage{amssymb} % For checkboxes
\usepackage{multirow}
\usepackage{algpseudocode}
\usepackage{xcolor}
\usepackage{colortbl}

\title{GA-S\textsuperscript{3}: Comprehensive Social Network Simulation with Group Agents}

\author{
 \textbf{Yunyao Zhang\textsuperscript{1}},
 \textbf{Zikai Song\textsuperscript{1}\thanks{Corresponding author}},
 \textbf{Hang Zhou\textsuperscript{1}},
 \textbf{Wenfeng Ren\textsuperscript{2}},
\\
 \textbf{Yi-Ping Phoebe Chen\textsuperscript{3}},
 \textbf{Junqing Yu\textsuperscript{1}},
 \textbf{Wei Yang\textsuperscript{1}}
\\
\\
 \textsuperscript{1}Huazhong University of Science and Technology,
 \textsuperscript{2}Silicon Universe Technology,
 \textsuperscript{3}La Trobe University
\\
 \small{
   \{ikostar, skyesong, henrrryzh, yjqing, weiyangcs\}@hust.edu.cn,
   ivenree@gjww.ai,
   phoebe.chen@latrobe.edu.au
 }
}

\begin{document}
\maketitle
\begin{abstract}
Social network simulation is developed to provide a comprehensive understanding of social networks in the real world, which can be leveraged for a wide range of applications such as group behavior emergence, policy optimization, and business strategy development.
However, billions of individuals and their evolving interactions involved in social networks pose challenges in accurately reflecting real-world complexities.
In this study, we propose a comprehensive \textbf{S}ocial network \textbf{S}imulation \textbf{S}ystem (GA-S\textsuperscript{3}) that leverages newly designed \textbf{G}roup \textbf{A}gents to make intelligent decisions regarding various online events. 
Unlike other intelligent agents that represent an individual entity, our group agents model a collection of individuals exhibiting similar behaviors, facilitating the simulation of large-scale network phenomena with complex interactions at a manageable computational cost.
% 
%We define the lifecycle of our group agents through three stages: existence, decision-making, and behavior. Firstly, the system retrieves multi-level population information based on environmental events, organizes it into a hierarchical tree structure, and ultimately generates group agents representing populations with similar behaviors influenced by the events. Then, the agents make decisions based on their role characteristics and the propagation of states and memory. Finally, our agents emulate human actions on the Internet and conduct interactions among multiple groups.
% 
% We collected data on 30 popular online events that span education, sports, entertainment, politics, economics, society, and technology in 2024. 
% 
Additionally, we have constructed a social network benchmark from 2024 popular online events that contains fine-grained information on Internet traffic variations.
The experiment demonstrates that our approach is capable of achieving accurate and highly realistic prediction results.
Code is open at \href{https://github.com/AI4SS/GAS-3}{https://github.com/AI4SS/GAS-3}.
\end{abstract}

\section{Introduction}

Online social networks have emerged as primary platforms for social activities, where users engage in different interactive behaviors, including chatting, posting, and sharing content. Social network simulations\cite{diffusion-online-social-networks2017survey} create virtual representations of social networks, modeling the behaviors, relationships, and information flows among individuals or entities within these networks. The goal of social network simulation\cite{social-network-analysis2004development} is to analyze and predict the outcomes that emerge from these interactions, and can be leveraged for a wide range of applications such as group behavior emergence, phenomenon prediction, policy optimization, and business strategy development.

Traditional social network simulation systems predominantly adhere to discrete events\cite{traditional-simulation-discrete-events} or system dynamics\cite{traditional-simulation-system-dynamics, Systems-Based-Approach-2012} focusing on predicting variables rather than elucidating underlying mechanisms or causal relationships. They often underestimate the heterogeneity of social behavior and the dynamism of social structures. Recently, research\cite{Stanford-town-2023, s3-2023, Stanford1000agents-2024} tends on social networks simulation focus on employing the large language model (LLM)\cite{gpt-4-2023, PaLM-2023, longdi} based agents to simulate individual entities and their behaviors. However, with billions of users on social networks, it is impossible to model each user individually through LLM-based agents\cite{Tencent-10billion-2024, knowledge-boundary-socail-agent-2024, luo2025gui}. Additionally, existing social simulation systems suffer from poor scalability and are frequently designed for particular events, thereby failing to comprehensively cover news events on the Internet.

To address the challenges of high complexity arising from billions of individuals and the poor scalability of social network simulations, we propose a comprehensive social network simulation system (GA-S\textsuperscript{3}) based on the intelligent agents\textemdash group agents. Group agents are designed to represent collections of individuals with similar online behaviors, rather than simulating each individual entity, thereby enabling the modeling of complex interactions at a manageable computational cost. Moreover, our group agents can adaptively generate user profiles based on network events, ensuring scalability across diverse online environments.

Our group agents consist of three primary modules: hierarchical generation, decision-reasoning, and action. These modules correspond to the three stages of the agents' life-cycle: existence, decision-making, and behavior. 
In the \textbf{hierarchical generation}, group agents' profiles are created based on the environmental event, including population, identity and characteristics. Groups are progressively differentiated within a hierarchical tree structure until they accurately represent populations with similar behaviors.
In the \textbf{decision-reasoning}, agents make the decision based on their role and the propagation of states and memory. We incorporate fine-grained factors such as emotion fading and forgetting probability to ensure alignment with real-world dynamics.
In the \textbf{action}, our agents emulate human actions on the Internet and conduct interactions among multiple groups.

We developed a Social Network Benchmark that compiles popular online events from 2024 across various domains and includes detailed information on fine-grained network traffic variations. Experimental results demonstrate that our GA-S\textsuperscript{3} model achieves accurate and highly realistic predictions.

% To evaluate the effectiveness of our GA-S\textsuperscript{3}, we collected data on 30 popular online events spanning education, sports, entertainment, politics, economics, society, and technology from January to November 2024. The 

\section{Realted Work}
\subsection{Social Simulation System (S\textsuperscript{3})}
% Social simulations are powerful tools for studying social activities, enabling large-scale artificial experiments to explore the long-term impacts of individual and group behaviors within social network structures\cite{SIR-model2002spread,simulation-model-2003, NC-Social-Media-2022universality}. By visualizing spatiotemporal dynamics that are difficult to observe empirically, these simulations reveal profound connections between micro-level behavioral changes and dynamic social structures\cite{Social-Simulation-overview-2014, strength-weak-ties2007complex}. They play a crucial role in advancing social science theories by validating hypotheses and adopting formalized methodologies\cite{Formal-Modeling-Social-Science-2019}. However, traditional approaches, such as discrete event-based simulations\cite{traditional-simulation-discrete-events} and system dynamics models\cite{traditional-simulation-system-dynamics, Systems-Based-Approach-2012}, emphasize variable prediction over uncovering mechanisms or causal relationships, often overlooking behavioral heterogeneity and the complexity of social dynamics. To address these limitations, we propose an integrated framework that combines agent-based modeling with system dynamics.

Social simulations are powerful tools for studying social activities, enabling large-scale experiments to explore the impacts of individual and group behaviors in social networks \cite{SIR-model2002spread, simulation-model-2003, NC-Social-Media-2022universality}. They reveal connections between micro-level behavioral changes and dynamic social structures\cite{Social-Simulation-overview-2014, strength-weak-ties2007complex} by visualizing spatiotemporal dynamics that are hard to observe empirically. PSP\cite{PSP-2018} identifies patterns of popularity stages in social media and uses pattern-matching techniques to predict future trends. While these simulations advance social science theories by validating hypotheses and adopting formalized methodologies\cite{Formal-Modeling-Social-Science-2019}, traditional approaches, such as event-based simulations\cite{traditional-simulation-discrete-events} and system dynamics models\cite{traditional-simulation-system-dynamics, Systems-Based-Approach-2012}, often prioritize prediction over uncovering causal mechanisms, neglecting behavioral heterogeneity. To address these limitations, we propose an integrated framework combining agent-based modeling with system dynamics.

\subsection{S\textsuperscript{3} based on Traditional Agent}
% Agent-based social simulations\cite{Old-Agent-Based-Social-Simulation-2002} utilize agent technology to study social phenomena by simulating the behaviors and interactions of individuals or groups. Each agent represents an individual or entity characterized by unique attributes, behavioral rules, and decision-making processes. By analyzing agent interactions, researchers can examine social phenomena, uncover behavioral patterns, and explore system dynamics.

% Schelling\cite{first-agent-based-model-schelling-1971, Micromotives-Macrobehavior-2006} pioneered the first agent-based virtual society simulation system, where agents represented individuals interacting within social processes. Using the concept of cellular automata\cite{Cellular-Automata-1998}, he examined housing segregation patterns\cite{Agent-Based-Modeling-2005}. Epstein and Axtell\cite{first-large-scale-agent-model-1996} advanced this approach with Sugarscape, the first large-scale agent-based model, expanding the scope from individual behaviors to entire societies. Grossi\cite{Multiagent-Systems2005} investigated organizational structures in multi-agent systems\cite{Multi-Agent-Systems-application-2018}, emphasizing their value and impact on the agents involved. Nevertheless, traditional agent-based social simulations continue to face challenges in addressing complexity, adaptability, and behavioral realism.

Agent-based social simulations \cite{Old-Agent-Based-Social-Simulation-2002, friendship-networks2009agent} utilize agent technology to model social phenomena by simulating the behaviors and interactions of individuals or groups, with each agent representing unique attributes, behavioral rules, and decision-making processes. Through the analysis of these interactions, researchers can identify behavioral patterns and explore system dynamics. Schelling \cite{first-agent-based-model-schelling-1971, Micromotives-Macrobehavior-2006} pioneered the first agent-based virtual society simulation, employing cellular automata \cite{Cellular-Automata-1998} to investigate housing segregation \cite{Agent-Based-Modeling-2005, segregation--social-networks2011emergence}. Epstein and Axtell’s Sugarscape \cite{first-large-scale-agent-model-1996} extended this methodology to model entire societies, while Grossi \cite{Multiagent-Systems2005} examined organizational structures and their influence within multi-agent systems \cite{Multi-Agent-Systems-application-2018}. Despite these advancements, traditional agent-based simulations continue to face challenges related to complexity, adaptability, and behavioral realism.

\subsection{S\textsuperscript{3} based on LLM Agent}

Advances in AI~\cite{luo2024deem, hu2025sf2t, song2025temporal} and multi-model technologies~\cite{li2024coupled, altrack, zhang2023depth} have enabled agent-based social simulations powered by large language models (LLMs)\cite{LLMs-generate-social-networks2024llms} to address complex challenges with robust comprehension and adaptability. JS Park et al.\cite{Stanford-town-2023} introduced generative agents simulating realistic behaviors\cite{model-credible-human-behaviors-2021, Few-Shot-Learners-LLM2020} in a simulated town, while S\textsuperscript{3}\cite{s3-2023} leveraged LLMs to build virtual social networks with advanced perception and reasoning. Despite these advancements, modeling every individual in real-world scenarios remains impractical due to the vast number of individuals and the complexity of behaviors. Existing approaches\cite{Stanford-town-2023, s3-2023, knowledge-boundary-socail-agent-2024, scaling-mutil-agents-graph-2024} are often manual, static, and lack scalability. To address these limitations, we propose a group-agents social network system powered by LLMs. This system integrates dynamic modeling to capture societal heterogeneity and evolving social structures, while automating agents generation based on environmental perception to enable efficient and scalable simulations.

\begin{figure*}[t]
\centering
\includegraphics[width=0.9\textwidth]{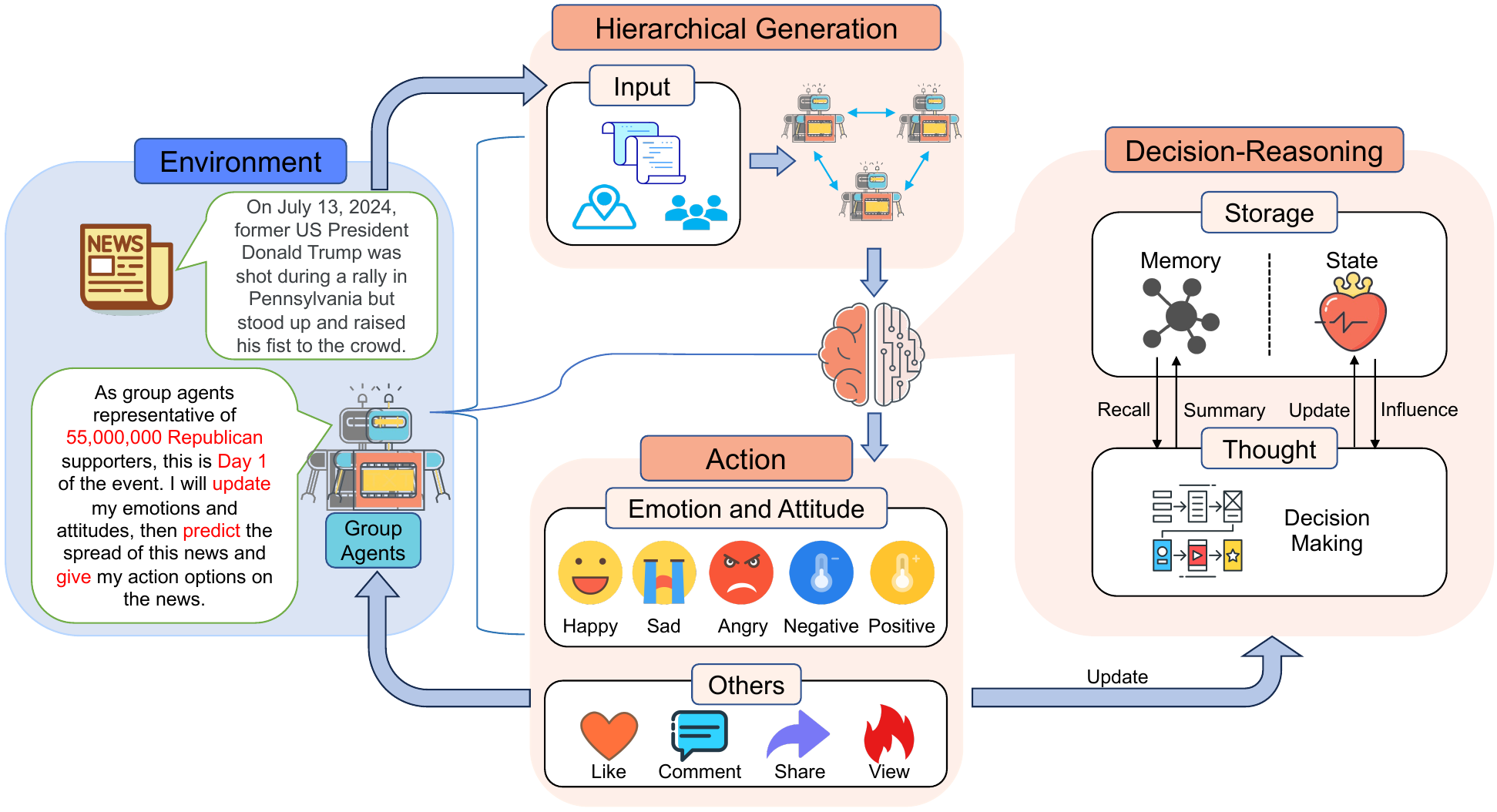}
\caption{Overview of our social network simulation system. It consists of three components: hierarchical generation, decision-reasoning, and action. Group agents are first generated through hierarchical generation to form perception of the environment and content. The perception and role attributes are stored in the memory and state of the decision-reasoning, respectively, and are leveraged to make decisions and actions.}
\label{fig1}
\end{figure*}

\section{Methodology}
Our comprehensive social network simulation system (GA-S\textsuperscript{3}) is shown in Figure \ref{fig1}, the group agents comprise three key modules: hierarchical generation, decision-reasoning, and action. 
The \textbf{hierarchical generation} is designed to automatically construct group agents based on the given network environmental events, shaping their perception from both the environment and content.
The \textbf{decision-reasoning}, functioning as the agent’s core and akin to a human brain, simulates human-like decision-making by reasoning about perceived content in alignment with role attributes.
The \textbf{action} simulates human behavior and enables the interaction with network entities or other agents according to decisions.

\begin{figure*}[t]
\centering
\includegraphics[width=0.99\textwidth]{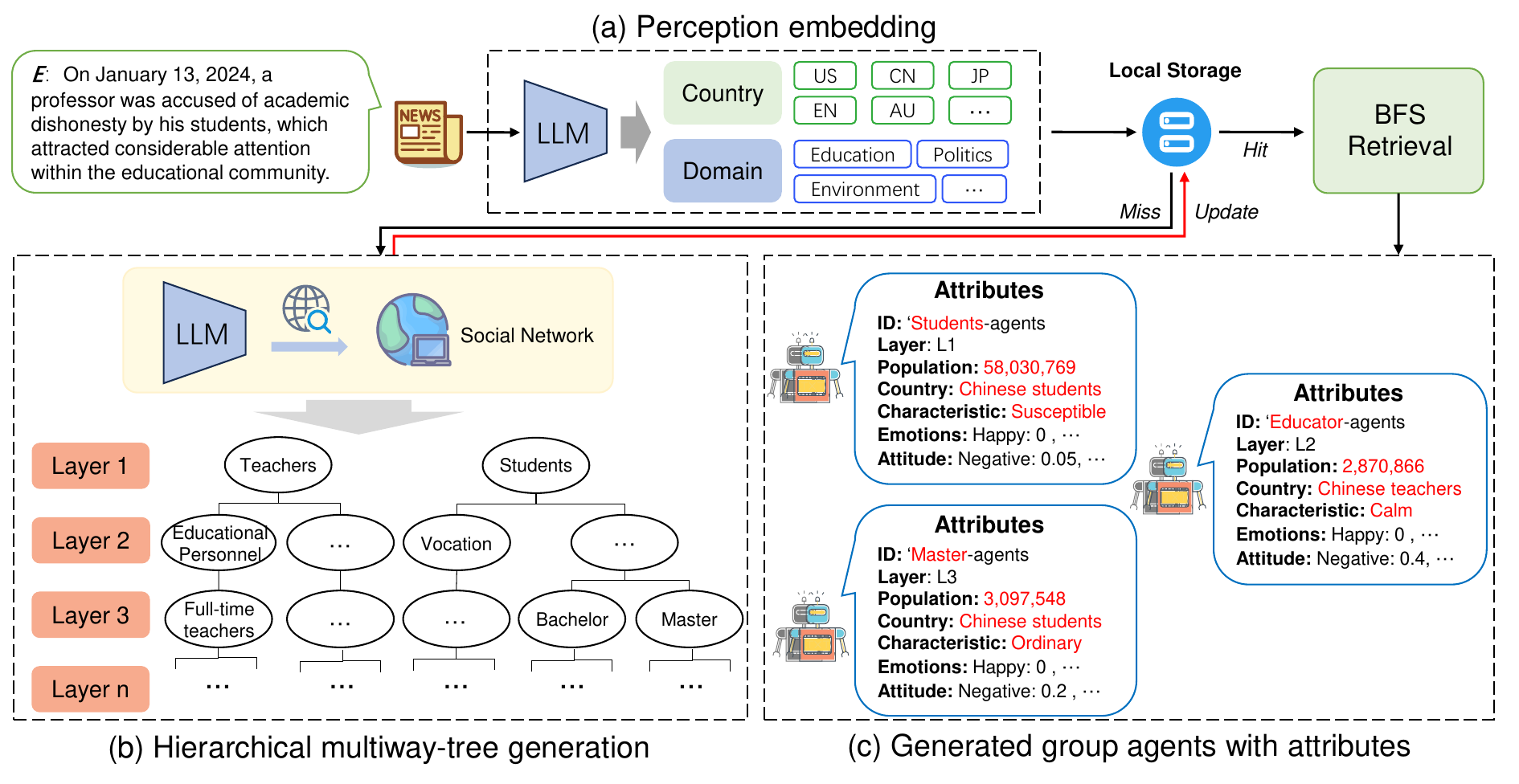}
\caption{The illustration for the hierarchical generation of group agents. (a) Perception embedding module that act as the "eyes" to perceive the environment; (b) hierarchical multiway-tree module is designed to generate group agents; (c) attributes show the profile of generated agents. We use an online educational event as an example.}
\label{fig2}
\end{figure*}

\subsection{Hierarchical Generation of Group Agents}
\label{section:3.1}
This section introduces the hierarchical generation process of group agents. The Perception Embedding subsection describes how agents perceive and respond to environmental events. The Hierarchical Multiway-tree Generation outlines the approach we developed to differentiate groups into progressively finer divisions, enabling them to represent similar behaviors. Finally, the Attributes subsection outlines the key attributes of the generated agents.

\subsubsection{Perception Embedding} 
Perception $\textcolor{blue}{O}$ embedding acts as the "eyes" of the group agents. When new environmental event $\textbf{\textit{E}}$ emerges in the virtual social network, the content of the event, along with its domain $\textbf{\textit{D}}$(e.g., education, politics, business, etc., based on widely accepted standards in media CNN and academic research) and country $\textbf{\textit{C}}$ as identified by a large language model, is stored in the agent's memory $\textcolor{orange!80}{M}$. This information forms the foundational perception of the environment and content, influencing subsequent reasoning processes and interactions. These foundational perceptions are continuously updated as new information becomes available.

\subsubsection{Hierarchical Multiway-tree Generation} 

To comprehensively generate group agents with similar behaviors, we adopt a top-down hierarchical approach, organizing populations into a multiway-tree structure, as shown in Figure \ref{fig2}. Each layer’s group agents are fine-grained divisions of the previous layer, generated using prompt engineering with large language models based on background information and environmental requirements. We also incorporate the Retrieval-Augmented Generation (RAG) technique for adaptive agents generation. Specifically, upon receiving an online event $\textbf{\textit{E}}$, the system analyzes its content to determine the domain $\textbf{\textit{D}}$ and country $\textbf{\textit{C}}$. It then leverages a large language model with network search capabilities to retrieve population information corresponding to the identified country and domain (Appendix Table \ref{tab7:group_find}), storing the data locally as a document for further modification if needed. This data is then integrated into the local hierarchical knowledge graph $\mathcal{G}$ (Figure \ref{fig2}(b)). \textbf{\textit{If events share the same country and domain, they correspond to the same group agents; differing countries or domains will result in different group agents}}. For future events, if corresponding country and domain information already exists, it is directly retrieved from $Layer_n$ of the knowledge graph using a Breadth-First Search (BFS)\cite{BFS-1984} algorithm, ensuring efficient traversal. Finally, the large language model (LLM) generates $n$ group agents based on the information in $Layer_n$ (Appendix Table \ref{tab8:group_agent_generation}).

% % 定义 tcolorbox 样式
% \tcbset{
%     colback=gray!10,    % 背景颜色（浅灰）
%     colframe=black,     % 边框颜色（黑色）
%     width=\textwidth,   % 宽度为页面宽度
%     boxrule=0.3mm,      % 边框线条粗细
%     arc=4mm,            % 圆角
%     width=0.48\textwidth,     % 宽度为页面宽度
%     title=              % 移除标题
% }
% \begin{table}[htbp]\small
% \centering
% \begin{tcolorbox}
% 1. Identify groups mentioned in the document.\\2. Generate a unique agent for each group.\\
% 3. Follow this format for each agent: 
% \begin{tabbing}
% \hspace{1em} \= agent \= \{n\}: \= (nth agent) \\
% \hspace{1em} \= id: \= \{group\}-agents \\
% \hspace{1em} \= description:  \= \parbox[t]{0.7\textwidth}{Representing \{\textbf{number}\} \{\textbf{country}\} \{\textbf{group}\}, reflecting...} \\
% \hspace{1em} \= characteristic:  \= \{\textbf{susceptible/ordinary/calm}\}
% \end{tabbing}
% \end{tcolorbox}

% \caption{Group agents generation template}
% \label{tab:rebuttal-group_agent_generation}
% \end{table}

\subsubsection{Attributes} 
% The generated group agents include an ID, population weight, country, group characteristics, emotions and attitudes. The ID, country indicates the group identity. The population weight influences both the frequency and intensity of group interactions. Group characteristics, which are fine-grain factors for capturing real-world effects, impact emotional and attitudinal fluctuations. These characteristics are classified into “susceptible population,” “ordinary population,” and “calm population.”\cite{Skepticism-Acceptance-fakenews-attitude-2024} This classification is designed to capture the varying degrees of emotional responsiveness and stability observed in real-world social dynamics. The “susceptible population” represents individuals who are highly sensitive to external stimuli, leading to significant emotional and attitudinal fluctuations. The “ordinary population” exhibits moderate responsiveness, reflecting average behavioral patterns. The “calm population,” by contrast, consists of individuals who maintain emotional stability and exhibit limited fluctuations, serving as a stabilizing force in the network. The emotions include happy, sad, and angry, and the attitudes include positive and negative. The attributes of the group agents are stored in their states $\textcolor{red}{S}$. Combined with perception, these attributes shape subsequent reasoning and interaction processes and are continuously updated.
The generated group agents include an ID, population, country, characteristics, emotions, and attitudes. The ID and country serve as unique identifiers for each group, while the population determines the frequency and intensity of interactions. Characteristics, classified as “susceptible,” “ordinary,” and “calm”\cite{Skepticism-Acceptance-fakenews-attitude-2024}, capture real-world emotional responsiveness and stability. The “susceptible” group is highly sensitive to stimuli, showing significant fluctuations; the “ordinary” group displays moderate and typical behavior; and the “calm” group maintains emotional stability. Emotions, including happy, sad, and angry, are quantified using an index, where a higher value indicates greater emotional intensity. Attitudes are categorized as either positive or negative, reflecting the agent’s overall evaluation of events or information. These attributes, stored in the agents’ states $\textcolor{red}{S}$, combine with perception to shape reasoning and interactions, which are continuously updated.

\subsection{Decision-Reasoning of Group Agents} 
\label{section:3.2}
The decision-reasoning consists of two components: storage and thought, as illustrated in Figure \ref{fig1}. We employ a state and memory updating strategy in the storage, which can be propagated and updated in reasoning, assisting in the decision-making process. Furthermore, we apply a Markov Network-based approach in the thought process to expand the information horizon. We additionally incorporate fine-grained factors that closely mirror real-world effects, ensuring alignment with practical scenarios.

% \subsubsection{Memory and State Update}
% Considering that a population's attention to specific online events is short-lived and diminishes over time, we introduced a memory updating mechanism\cite{agent-memory-2024my, LLM-memory-surver-2024} into the system. Specifically, the agent's 1) perception, 2) decisions, and 3) actions are stored in a queue structure, where newer information replaces older ones as the social network evolves, leading to gradual forgetting. This memory system not only reflects the agent's perception of the environment and information but also records its interactions, thereby shaping its subsequent reasoning processes.

% We collectively define the attributes of group agents as their state $\textcolor{red}{S}$. Here, $\mathcal{E}$ specifically represents the combination of emotions and attitudes in the agent's attributes. This state is initialized during the group agents generation process and dynamically updated in response to new online events, with $\mathcal{E}$ being triggered and adjusted accordingly. Our group agents, representing distinct groups of people \cite{emotional-reactions-different-groups-2005}, can exhibit consistent emotional and attitudinal patterns resembling those of real individuals.

\subsubsection{State and Memory of Storage}
The storage system comprises two core components: States $\textcolor{red}{S}$, and Memory $\textcolor{orange!80}{M}$.  

The State $\textcolor{red}{S}$ encapsulates the attributes of group agents. Here, $\mathcal{E}$ specifically represents the combination of emotions and attitudes within the agents' attributes, simulating the emotional and attitudinal patterns observed in real human groups\cite{emotional-reactions-different-groups-2005}. 

The Memory $\textcolor{orange!80}{M}$ is based on a queue updating mechanism\cite{agent-memory-2024my, LLM-memory-surver-2024}, reflecting the fact that people’s attention to specific online events is short-lived and diminishes over time. Specifically, the agent’s 1) perception, 2) decisions, and 3) actions are stored in a queue structure, where newer information replaces older data as the social network evolves, leading to gradual forgetting. %This memory system not only reflects the agent's perception of the environment and information but also records its interactions, thereby shaping its subsequent reasoning processes. 

\subsubsection{Markov Network-based Reasoning}
\label{sec:3.2.2 markov network}

To capture the evolution of an agent's state and memory in decision-making, we employed a Markov Network framework\cite{Markov-network-with-decision-trees2010}. The state transition is defined as follows:
\begin{equation}
\begin{split}
P(\textcolor{red}{S}_i^t | \textcolor{red}{S}_{i}^{t-1}, \mathcal{E}_i^t, \textcolor{orange!80}{M}_i^{t-1}) =  \alpha_1 P(\textcolor{red}{S}_i^{t-1}) \\+ \alpha_2 P(\mathcal{E}_i^t) + \alpha_3 P(\textcolor{orange!80}{M}_i^{t-1}),
\end{split}
\end{equation}
where $P(\textcolor{red}{S}_i^t)$ represents the probability distribution over the state of agent $i$ at time $t$, and $\alpha$ represents fading parameters controlling attribute decay rates. 

At each time step, the emotional response $\mathcal{E}_i^t$ is updated based on the agent's perception $\textcolor{blue}{O}_i^t$, and its prior state $\textcolor{red}{S}_i^{t-1}$ as follows:
\begin{equation}
P(\mathcal{E}_i^t | \textcolor{blue}{O}_i^t, \textcolor{red}{S}_i^{t-1}) = \text{LLM}(\textcolor{blue}{O}_i^t, \textcolor{red}{S}_i^{t-1}),
\end{equation}
where $\textcolor{blue}{O}_i^t$ includes the agent's perception influenced by its own environment $E_i^t$, as determined by:
\begin{equation}
\textcolor{blue}{O}_i^t \gets \text{Perception}(E_i^t).
\end{equation}

The decision-making process is refined by an LLM policy $\pi$, which determines the subsequent action $A_i^t$ based on the updated state $\textcolor{red}{S}_i^t$:
\begin{equation}
\begin{split}
P(A_i^t | \textcolor{red}{S}_i^t) = \pi(\textcolor{red}{S}_i^t),
\end{split}
\end{equation}

Finally, the memory $\textcolor{orange!80}{M}_i^t$ is updated to reflect its accumulation of prior perception and actions:
\begin{equation}
P(\textcolor{orange!80}{M}_i^t | \textcolor{orange!80}{M}_i^{t-1}, A_i^t, \textcolor{blue}{O}_i^t) = \textcolor{orange!80}{M}_i^{t-1} + A_i^t + \textcolor{blue}{O}_i^t,
\end{equation}

The entire framework is managed through four managers: the event manager, memory manager, state manager, and object manager, where $\textcolor{blue}{O}$, $\textcolor{red}{S}$, and $\textcolor{orange!80}{M}$ are \textit{open text} and stored in \texttt{JSON} format. This framework enables a dynamic and iterative process where the agent’s state, actions, and memory evolve adaptively in response to environmental changes and inter-agent communication. Furthermore, by utilizing this framework, the bias in the responses of LLM agents can be reduced, thus improving the repeatability and consistency of the experimental process.

\subsubsection{Fine-grain factors for real-world effect}
To enhance the resemblance of the virtual social network to a real-world social network, we introduced four fine-grain factors: 1) population weight, 2) group characteristics, 3) emotion fading, and 4) forgetting probability. Population weight, derived from real data, influences the activity levels and interaction frequencies of groups. Group characteristics determine the amplitude of emotional and attitudinal fluctuations, with susceptible groups experiencing the greatest fluctuations, followed by ordinary groups, and calm groups experiencing the least. The emotion fading represents the gradual decline of emotions and attitudes over time. The forgetting probability affects short-term memory\cite{long-short-term-memory-2012}, leading to the gradual fading of memories related to past perception and events.

\begin{table*}[h!]\normalsize
\centering
\caption{Ablation experiments. Default settings are marked in gray. See Section \ref{sec:ablation} for details.}
\label{tab:ablation1}
\begin{tabular}{cccc|ccccc}
\toprule
\multirow{2}*{\#} & \multirow{2}*{Layer} & \multirow{2}*{Memory} & \multirow{2}*{State} & \multirow{2}*{t-test~$\downarrow$} & \multirow{2}*{MAPE~$\downarrow$} & \multicolumn{2}{c}{DTW}& \multirow{2}*{Z-score} \\
\cmidrule(lr){7-8}
&&&&&& Mean~$\downarrow$ & Std~$\downarrow$ & \\
\midrule
1&L1&\checkmark&\checkmark& 0.829 & 68.78\% & 3.38e+07 & 0.4186 & 0.84 \\
2&L2&\checkmark&\checkmark& 0.603 & 33.73\% & 2.84e+07 & 0.3927 & 0.21 \\
3&L3&-&-& 2.212 & 2884.25\% & 7.80e+08 & 1.0263 & 2.32 \\
4&L3&\checkmark&-& 2.189 & 1338.74\% & 1.39e+08 & 0.7653 & 1.41 \\
5&L3&-&\checkmark& 1.986 & 400.71\% & 8.78e+07 & 0.6674 & 0.25 \\
\rowcolor{gray!30}
6&L3&\checkmark&\checkmark& 0.389 & 16.48\% & 1.30e+07 & 0.1890 & 0.81 \\
\bottomrule
\end{tabular}
\end{table*}

\subsection{Action of Group Agents}
Actions enable agents to simulate human behavior online and facilitate interactions between multiple groups. Users can perform the following actions on a event item: view, like, comment, share and predict. Viewing involves reading the event, liking indicates interest, commenting expresses thoughts, sharing distributes the event, and predicting\cite{EMNLLP-predict-political-2024, Electionsim-twitter-2024} involves projecting future outcomes based on the content. We hypothesize that viewing is the primary action, with views significantly outnumbering other actions. Interactions from fake users are excluded as they do not represent real populations. Real user groups first view the event and its related information (e.g., view counts, likes, comments, shares), which generates emotions and attitudes, driving further interactions and new content creation. For instance, 10,000 students might view an event about academic dishonesty, with 2,000 liking it and 500 sharing or commenting, leading to updates in event information and influencing subsequent interactions.

\begin{table}[h!]\small
\centering
\caption{Comparisons with other social simulation systems on our SNB benchmark}
\label{tab:all}
\begin{tabular}{@{ }lcccc@{}}
\toprule
\multirow{2}*{Method} & \multirow{2}*{t-test~$\downarrow$} & \multirow{2}*{MAPE~$\downarrow$} & \multicolumn{2}{c}{DTW}\\  
\cmidrule(lr){4-5}
&&&  Mean~$\downarrow$ & Std~$\downarrow$ \\
\midrule
PSP & 1.310 & 69.12\% & 3.40e+07 & 0.4207 \\
S$^3$ & 1.820 & 68.66\% & 3.09e+07 & 0.4035 \\
\midrule
GA-S\textsuperscript{3}(ours) & \textbf{0.389} &  \textbf{16.48\%} & \textbf{1.30e+07} & \textbf{0.1890}\\
\bottomrule
\end{tabular}
\end{table}

% \begin{table*}[h!]\footnotesize  % 开始一个双栏宽度的表格环境，尝试将表格放置在此处
% \centering          % 将表格居中对齐
% \caption{Comparison of multi-layer group agents performance with other methods}  % 表格标题
% \label{tab1:effectiveness}  % 表格标签，用于引用
% \begin{tabular}{l c c c c c c c c}  % 定义表格列格式：两列左对齐，七列右对齐
% \toprule  % 顶部粗线
% Method & \multicolumn{4}{c}{t-test} & \makecell{MAPE\\(Views)} & \multicolumn{2}{c}{DTW} & Z-score \\  % 主标题行
% \cmidrule(lr){2-5} \cmidrule(lr){7-8}  % Answering 和 User 部分的细线
% & Views & Likes & Comments & Shares & & Mean & \makecell{Std\\(Normalized)} \\  % 次级标题行
% \midrule  % 数据与标题之间的中粗线
% PSP & -1.989 & -1.785 & -1.118 & -0.349 & 69.12\% & 3.40e+07 & 0.4207 & -\\  % 数据行
% S$^3$ & -1.026 & -1.094 & -1.966 & -3.192 & 68.66\% & 3.09e+07 & 0.4035 & -1.89\\  %

% \midrule  % 分组数据之间的中粗线
% Ours-layer1 & -0.785 & 1.218  & \textbf{-0.083} & -1.229 & 68.78\% & 3.38e+07 & 0.4186 & 0.84\\
% Ours-layer2 & 0.597  & 0.951  & 0.640  & -0.223 & 33.73\% & 2.84e+07 & 0.3927 & 0.21\\
% Ours-layer3 & \textbf{0.292}  & \textbf{0.608}  & 0.471  & \textbf{-0.183} & \textbf{16.48\%} & \textbf{1.30e+07} & \textbf{0.1890} & 0.81\\
% \midrule  % 分组数据之间的中粗线
% w/o Memory & 2.728 & 1.732  & 1.687 & 1.796 & 400.71\% & 8.78e+07 & 0.6674 & 0.25 \\
% w/o State & 2.175 & 2.299  & 1.892 & 2.390 & 1338.74\% & 1.39e+08 & 0.7653 & 1.41\\
% \bottomrule  % 底部粗线
% \end{tabular}
% \end{table*}

\section{Experiment}
\subsection{Benchmark}
Due to the absence of temporal variations on network traffic in existing datasets, we have developed a \textbf{S}ocial \textbf{N}etwork \textbf{B}enchmark (SNB), which incorporates detailed information on fine-grained network traffic variations.

\noindent\textbf{Data collection.}
The datasets in our SNB have been collected on the 2024 popular online events from three major platforms: Twitter (referred to as Platform X), Reddit, and Weibo. Twitter\cite{twitter-2010} is a widely used microblogging platform known for its real-time news and social interactions; Reddit\cite{reddit-2015} is a community-driven platform that hosts discussions on a wide range of topics; Weibo\cite{weibo-2014} is a popular Chinese social media platform.

\noindent\textbf{Statistics.}
Our SNB dataset consists of 30 online events, each containing a title, content, metadata, and network traffic variations over a 7-day period. These events cover a wide range of domains, including education, politics, business, technology, culture, sports, health, entertainment, environment, and economy. They originate from various countries, such as China, the United States, and Japan, representing different levels of popularity.

\noindent\textbf{Evaluation Metrics.} To comprehensively evaluate the performance of the system, we selected multiple metrics from various perspectives.
\begin{itemize}[itemsep=0.1em,topsep=0.1em]
    \item \textbf{t-test} \cite{paired-t-test-2014} is employed to assess whether there is a significant difference between the means of two groups. 
    \item \textbf{Mean Absolute Percentage Error (MAPE)}\cite{mape-1999} provides a clear, percentage-based measure of error, making it useful for evaluating prediction accuracy in relative terms. 
    \item \textbf{Dynamic Time Warping (DTW)}~\cite{DTW-2007} is highly effective for comparing time series, even when the sequences vary in speed or timing. In our context, DTW is employed to evaluate the model's ability to capture the \textbf{complexity and dynamics} of real-world social interactions. Specifically, we compute the alignment between predicted and observed \textit{Network Traffic Variations} during event time periods, as a proxy for behavioral realism. A detailed explanation of the DTW computation process is provided in Appendix~\ref{DTW}.
    \item \textbf{Z-score} \cite{z-score-2016} is used to evaluate errors in repeated results, with deviations classified as small if the absolute value of the Z-score is less than \textbf{1}. 
\end{itemize}

\begin{table}[h!]\footnotesize
\centering
\caption{Information of selected three events within the educational domain}
\label{tab:event}
\begin{tabular}{@{}cll@{}}
\toprule
Event & Descriptions & Distinctions\\ 
\midrule
\#2 & \makecell[l]{A school cafeteria \\ where was reported \\ spoiled pork} & \makecell[l]{An explosive event \\ with a flood of traffic \\ and two peaks}\\
\midrule
\#7 & \makecell[l]{An academic dishonesty \\ where a professor \\ was dismissed} & \makecell[l]{A hot event with \\ a viewership peak \\ on the 3rd day}\\
\midrule
\#14 & \makecell[l]{1,477 freshmen forfeited \\ admission for failing to \\ enrollment on time} & \makecell[l]{A common event \\ with views peaking \\ on the 2nd day}\\
\bottomrule
\end{tabular}
\end{table}

\begin{figure}[ht]
    \centering
    \includegraphics[width=\columnwidth]{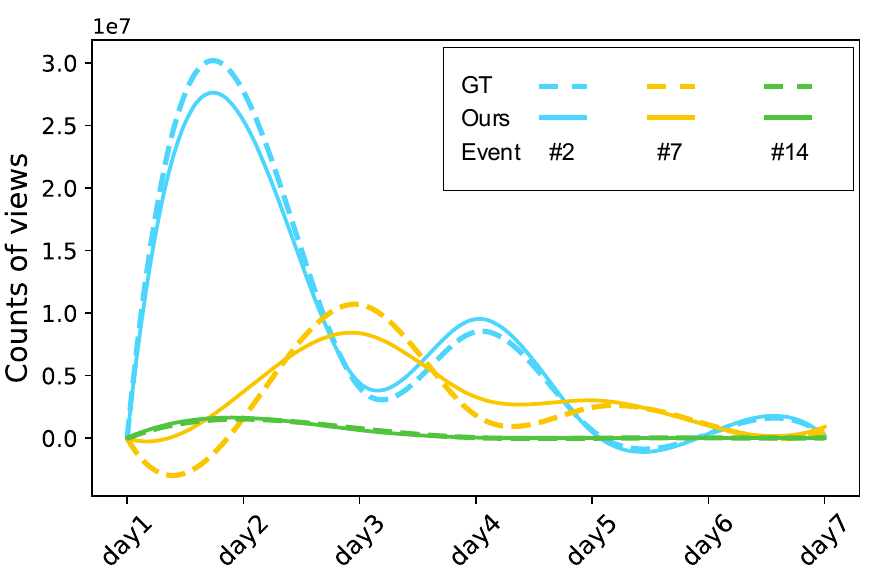}
    \caption{Network traffic trends in three events. We use the counts of views as a proxy for network traffic.}
    \label{fig5:trend-comparison}
\end{figure}

\subsection{LLM Setup}
Our group agents, as introduced in Section~\ref{section:3.2}, are powered by the open-source LLM LLaMA3-8B~\cite{llama3-2024}, chosen for its transparency, the flexibility of local deployment, and its support for deterministic outputs with a low temperature setting (\texttt{temperature=0.1}). Importantly, our method does not require any fine-tuning or task-specific adaptation. The model can be deployed directly, enabling fast and low-cost integration in practical scenarios.

In contrast, the Hierarchical Multiway Tree-designed Structure described in Section~\ref{section:3.1} leverages Kimi’s large model~\cite{kimi-report-2024} for its internet search capabilities. Due to potential noise in the retrieved content, we additionally integrate GPT-4~\cite{gpt-4-2023} to perform advanced data cleaning, which improves the overall accuracy and reliability of the system.

\subsection{Comparison}
We need to emphasize that there are limited solutions for simulating social networks, particularly for billions of individuals in Internet events. However, for comparative evaluation, we selected and re-implemented the PSP\cite{PSP-2018} and the S\textsuperscript{3}\cite{s3-2023} on our SNB benchmark. These methods support social simulation using model-based and agent-based approaches, respectively.
% trends and patterns
PSP focuses on predicting trends of popularity stages at both micro and macro levels, while S\textsuperscript{3} employs 1,000 fixed agents to predict news events, scaling the results to the population benchmark for broader predictions. As shown in Table~\ref{tab:all}, our GA-S\textsuperscript{3} method significantly outperforms both PSP (model-based) and S\textsuperscript{3} (agent-based) across all evaluation metrics, including t-test, MAPE, and the mean and standard deviation of DTW.

\begin{table*}[ht]\small
\centering
\caption{Detailed information on view counts for 3 specific events. All values are expressed in units of 10,000 (10k)}
\label{tab2:analysis}
\begin{tabular}{cccc@{ }c|c@{ }cc@{ }cc@{ }c}
\toprule
Layer & Agent & Character & Population & (\%) & \multicolumn{2}{c}{Event \#2} & \multicolumn{2}{c}{Event \#7} &\multicolumn{2}{c}{Event \#14} \\ 
\midrule  % 数据与标题之间的中粗线
\multirow{3}*{L1} & Students & susceptible &5803&(94.4\%) &4319&(93.1\%)&1239&(91.6\%)&360&(90.6\%)\\
& Teachers & calm & 345 & (5.6\%) &  319 & (6.9\%) & 113 & (8.4\%) & 38 & (9.4\%)\\
\cmidrule(lr){2-11}
& \multicolumn{2}{c}{Total (2 agents)} & 6148 & & 4638 &  & 1352 &  & 398 &\\
\midrule
\multirow{4}*{L2} & Vocation & susceptible & 3472 & (56.5\%) & 2508 & (57.2\%) &  1141 & (56.8\%) & 195 & (55.8\%)\\
& Educators & calm & 287 & (4.7\%) & 206 & (4.7\%) & 104 & (5.2\%) & 16 & (4.6\%) \\
& ... &  &  &  &&&&&&\\
\cmidrule(lr){2-11}
& \multicolumn{2}{c}{Total (5 agents)} & 6148 & & 4385 &  & 2010 &  & 350 &\\
\midrule
\multirow{5}*{L3} & Bachelor & susceptible & 1966 & (32\%) & 1272 & (31\%) & 710 & (36.2\%) & 81 & (36.3\%) \\
& Master & ordinary & 310 & (5\%) & 180 & (4.4\%) & 77 & (3.9\%) & 9 & (4.1\%)\\
& FullTime & calm & 201 & (3.3\%) & 135 & (3.3\%) & 41 & (2.1\%) & 5 & (2.3\%)\\
& ... &  &  &  &&&&&&\\
\cmidrule(lr){2-11}
& \multicolumn{2}{c}{Total (16 agents)} & 6148 & & 4170 &  & 1963 &  & 224 &\\
\midrule
\multicolumn{4}{c}{Ground-truth} && 4141 &  & 1817 &  & 234 &\\
\bottomrule  % 底部粗线
\end{tabular}
\end{table*}

\begin{table}[h!]\normalsize
\centering
\caption{Distribution of actions under t-test score.}
\label{tab:ablation2}
\begin{tabular}{@{ }l|cccc@{}}
\toprule
Layers & Views & Likes & Comments & Shares\\ 
\midrule
L1 & -0.785 & 1.218  & \textbf{-0.083} & -1.229\\
L2 & 0.597  & 0.951  & 0.640  & -0.223 \\
L3 & \textbf{0.292}  & \textbf{0.608}  & 0.471  & \textbf{-0.183}\\
\bottomrule
\end{tabular}
\end{table}

\begin{figure}[htb]
\centering
\subfloat[Emotion trends]{\includegraphics[width=0.5\linewidth]{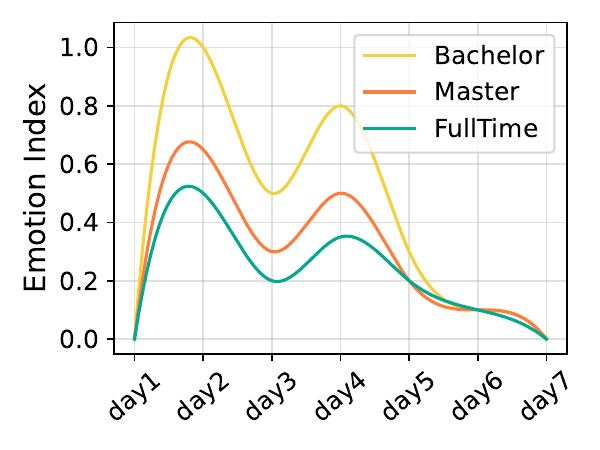}}
\hfill
\subfloat[Attitude trends]{\includegraphics[width=0.5\linewidth]{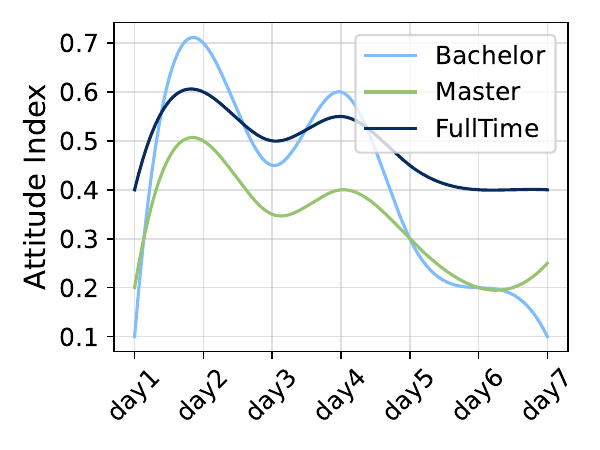}}

\subfloat[w/o fading\&forgetting ]{\includegraphics[width=0.5\linewidth]{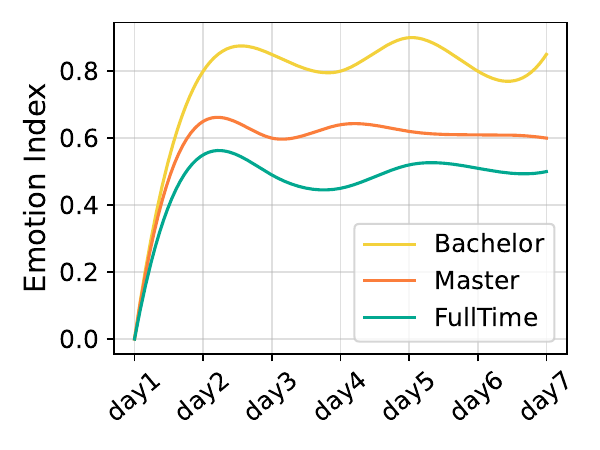}}
\hfill
\subfloat[w/o character]{\includegraphics[width=0.5\linewidth]{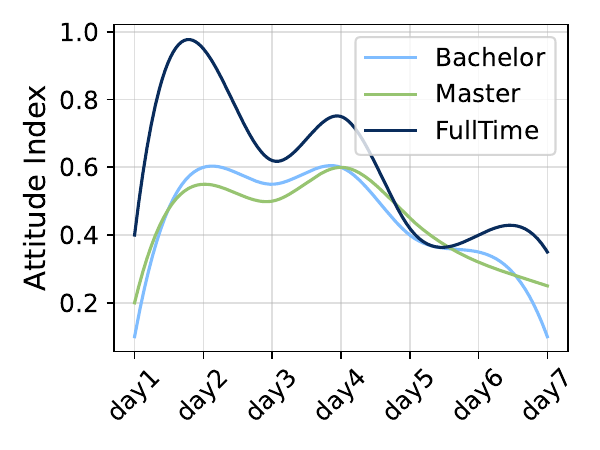}}
\caption{Emotion and attitude trends for each agent in the Event \#2.}
\label{fig6}
\end{figure}

\subsection{Ablation Study}
\label{sec:ablation}
% To evaluate the effectiveness of different components in GA-S$^3$, we conducted an ablation study by removing specific elements in the framework. The analysis included the following: 1) omitting the memory design, 2) excluding the state component, and 3) individually ablating the fine-grained factors of emotion fading and group characteristics.
% xxx(why select these 3 news)

\noindent\textbf{Reproducibility.}
We have evaluated the reproducibility by conducting over five tests under identical conditions and assessing the results using the Z-score. A Z-score below 3 is generally considered an acceptable deviation. As shown in Table~\ref{tab:ablation1} (line \#1, \#2, \#6), our Z-scores are consistently below 1, indicating excellent reproducibility.

\noindent\textbf{Hierarchical multiway-tree generation.} 
We have ablated our hierarchical design for generating group agents. As the hierarchy deepens, the accuracy of agents construction improves. As shown in Table \ref{tab:ablation1}, the experimental results confirm the effectiveness of our design, with the simulation accuracy of the L3 layer significantly exceeding that of the L1 and L2 layers.

\noindent\textbf{Memory and state in decision-reasoning.} 
As shown in Table \ref{tab:ablation1}, incorporating the memory module improves the t-test score from 1.986 to 0.389 (line \#5 $\rightarrow$ \#6), and the state module enhances the score from 2.189 to 0.389 (line \#4 $\rightarrow$ \#6). These results strongly validate the critical role of the memory and state update modules in our GA-S\textsuperscript{3}.

\noindent\textbf{Distribution of different actions.}
We have analyzed the results of four distinct online behaviors: views, likes, comments, and shares. The experimental results, presented in Table \ref{tab:ablation2} under t-test scores, reveal that while the number of views is significantly higher than the other three actions, the simulation results demonstrate that all four actions are modeled with high accuracy.

\noindent\textbf{Fine-grain factors.}
We have ablated the four fine-grain factors that are introduced for more suitable to real world including: population weight, group characteristics, emotion fading, and forgetting probability. 
\textbf{(1) For population weight}, we present the detailed information for 3 events in the same domain (Event \#2, \#7, \#14), as shown in Table~\ref{tab2:analysis}. The popularity of different agents plays a decisive role in determining the volume of network traffic, with the ratio of view counts closely aligning with the ratio of population;
\textbf{(2) For group characteristics (character)}, from Table~\ref{tab2:analysis}, we know that the character of full-time faculty agents (FullTime) is "calm". Comparing Figure \ref{fig6} (b) and (d), we observe that removing the character leads to a significant increase in Fulltime's attitude value, which does not align with the actual situation. This highlights the critical role of the character in influencing the online behavior of specific group roles.
\textbf{(3) For emotion fading and forgetting probability}, by comparing Figure \ref{fig6} (a) and (c), it is evident that the absence of emotion fading and forgetting probability causes significant distortion in network traffic prediction.

% \textbf{Layer1} & 2 group agents & & 2352 & 4638 & 724 & 1352 & 381 & 398 \\
% Students & 5803(94.4\%) & susceptible & 2100(89.3\%) & 4319(93.1\%) & 684(94.4\%) & 1239(91.6\%) & 350(91.9\%) & 360(90.6\%) \\  % 数据行
% Teachers & 345(5.6\%) & calm & 252(10.7\%) & 319(6.9\%) & 40(5.6\%) & 113(8.4\%) & 31(8.1\%) & 38(9.4\%) \\  % 数据行
% \midrule  % 分组数据之间的中粗线
% \textbf{Layer2} & 5 group agents & & 3030 & 4385 & 1173 & 2010 & 301 & 350 \\
% Vocation & 3472(56.5\%) & susceptible & 1833(60.5\%) & 2508(57.2\%) & 694(59.2\%) & 1141(56.8\%) & 163(54.2\%) & 195(55.8\%) \\  % 数据行
% Educators & 287(4.7\%) & calm & 99(3.2\%) & 206(4.7\%) & 44(3.7\%) & 104(5.2\%) & 13(4.3\%) & 16(4.6\%) \\  % 数据行
% \midrule  % 分组数据之间的中粗线
% \textbf{Layer3} & 16 group agents & & 2641 & 4170 & 1096 & 1963 & 153 & 224 \\
% Bachelor & 1966(32\%) & susceptible & 891(33.7\%) & 1272(31\%) & 393(35.9\%) & 710(36.2\%) & 53(34.7\%) & 81(36.3\%) \\  
% Master & 310(5\%) & ordinary & 134(5.1\%) & 180(4.4\%) & 40(3.6\%) & 77(3.9\%) & 6(4\%) & 9(4.1\%) \\  
% FullTime & 201(3.3\%) & calm & 95(3.6\%) & 135(3.3\%) & 23(2.1\%) & 41(2.1\%) & 3(2\%) & 5(2.3\%) \\ 

\subsection{Analysis}

\noindent\textbf{Behavioral diversity of fixed group agents.}
One potential phenomenon is that a specific agent may exhibit only a single pattern of online behavior, even across different Internet events. To verify this hypothesis, we selected three events from the same educational domain, each with significant distinctions, as shown in Table~\ref{tab:event}. We set the group agents in these three events to be identical. The prediction results, shown in the Figure~\ref{fig5:trend-comparison}, demonstrate that our agents can exhibit considerable diversity across different events, with prediction curve closely matching the ground-truth. This confirms that our agents are capable of treating different online events distinctively, which aligns well with real-world behavior.

\noindent\textbf{Relationship between emotion/attitude and network traffic.}
Figure~\ref{fig6}~(a) and~(b) illustrate the trends of emotions and attitudes for three different agents in Event \#2. Compared to the network traffic shown by the blue lines in Figure \ref{fig5:trend-comparison}, we observe there are two peaks in emotions and attitudes, which align with the traffic trends. However, while the second peak in emotions and attitudes is closer to the first peak, the gap between the two peaks in traffic trends is larger. This indicates that emotions/attitudes and traffic are not perfectly correlated, but only exhibit a limited degree of alignment.

\section{Conclusion}

In this study, we introduce a comprehensive social network simulation framework based on group agents. This framework effectively constructs virtual representations of social networks and simulates human behaviors. We have designed group agents capable of modeling a collective of individuals and engaging in social network interactions. The group agents incorporate hierarchical generation, decision-reasoning, and action modules corresponding to the three stages in the life-cycle of existence, decision, and behavior. This design enables the simulation of large-scale network phenomena with complex interactions while maintaining a manageable computational cost. Experiments conducted on our BNS benchmark demonstrate that our GA-S\textsuperscript{3} system delivers accurate and realistic predictive outcomes.

\section*{Limitation}
Our GA-S\textsuperscript{3} has demonstrated exceptional predictive accuracy; however, it is important to acknowledge the following limitations: 
(1) Limited reasoning capability. The current agent reasoning is directly produced by LLMs, which may restrict the depth of deliberation in complex scenarios. This limitation can potentially be addressed through techniques such as Chain-of-Thought\cite{COT-2022chain};
(2) The diversity of our SNB benchmark. Our SNB offers a key advantage in its fine-grained network traffic variations, complemented by efforts to include a wide range of news topics. Nevertheless, given the vast amount of information available online, there remains room to further enhance the dataset’s diversity. In addition, it is crucial to emphasize that our data sources fully comply with General Data Protection~\cite{GDPR-2017eu}; to safeguard user privacy, all individuals featured in publicly available social events have been anonymized;
(3) Lack of explicit network and interactivity. Although group-level dynamics induce implicit structures via domain and geographic boundaries, the absence of explicit connections and agent-level interactions may limit the emergence of more realistic social behaviors;
(4) Limited adaptability in group generation. The current group generation relies on a fixed hierarchical structure, which may not adapt well to diverse scenarios. Future improvements could explore dynamic layer adjustments to enhance adaptability.

\section*{Acknowledgments}
This work is supported by the National Key Research and Development Program of China (No.2020YBF2901202), National Natural Science Foundation of China (NSFC No. 62272184 and No. 62402189), the China Postdoctoral Science Foundation under Grant Number GZC20230894, the China Postdoctoral Science Foundation (Certificate Number: 2024M751012), and the Postdoctor Project of Hubei Province under Grant Number 2024HBBHCXB014. The computation is completed in the HPC Platform of Huazhong University of Science and Technology.

% This study developed an agent system based on an open-source large language model (LLM). Efforts were made to enhance the agents' personalized performance through attribute design and related methods. However, the accuracy of the agents in specialized domains was constrained by the lack of fine-tuning of the LLM for specific fields. Additionally, the interactive behavior of group agents remains limited, hindering their ability to autonomously generate new events that could trigger significant actions. In terms of dataset design, to protect user privacy, individuals in the dataset of publicly available social events were anonymized. While efforts were made to ensure dataset diversity, there remains potential for further expansion. Nevertheless, the system successfully analyzes events across multiple domains, providing a solid foundation for future development. Future research should focus on enhancing the expressiveness and generative capabilities of agents by fine-tuning the LLM for domain-specific applications and optimizing agent behavior. Moreover, expanding the dataset's coverage and diversity will improve the model's generalization capacity, increase simulation accuracy, and enhance verification reliability.

% Bibliography entries for the entire Anthology, followed by custom entries
%\bibliography{anthology,custom}
% Custom bibliography entries only

\bibliography{main}
\appendix

% 设置附录中的表格和图像编号为罗马数字
\renewcommand{\thetable}{\Roman{table}}
\renewcommand{\thefigure}{\Roman{figure}}
\setcounter{table}{0}
\setcounter{figure}{0}

% \begin{figure*}
% \centering

% \end{figure*}

\let\oldtwocolumn\twocolumn
\renewcommand\twocolumn[1][]{%
    \oldtwocolumn[{#1}{
\begin{center}
\vspace{-6em}
\includegraphics[width=\textwidth]{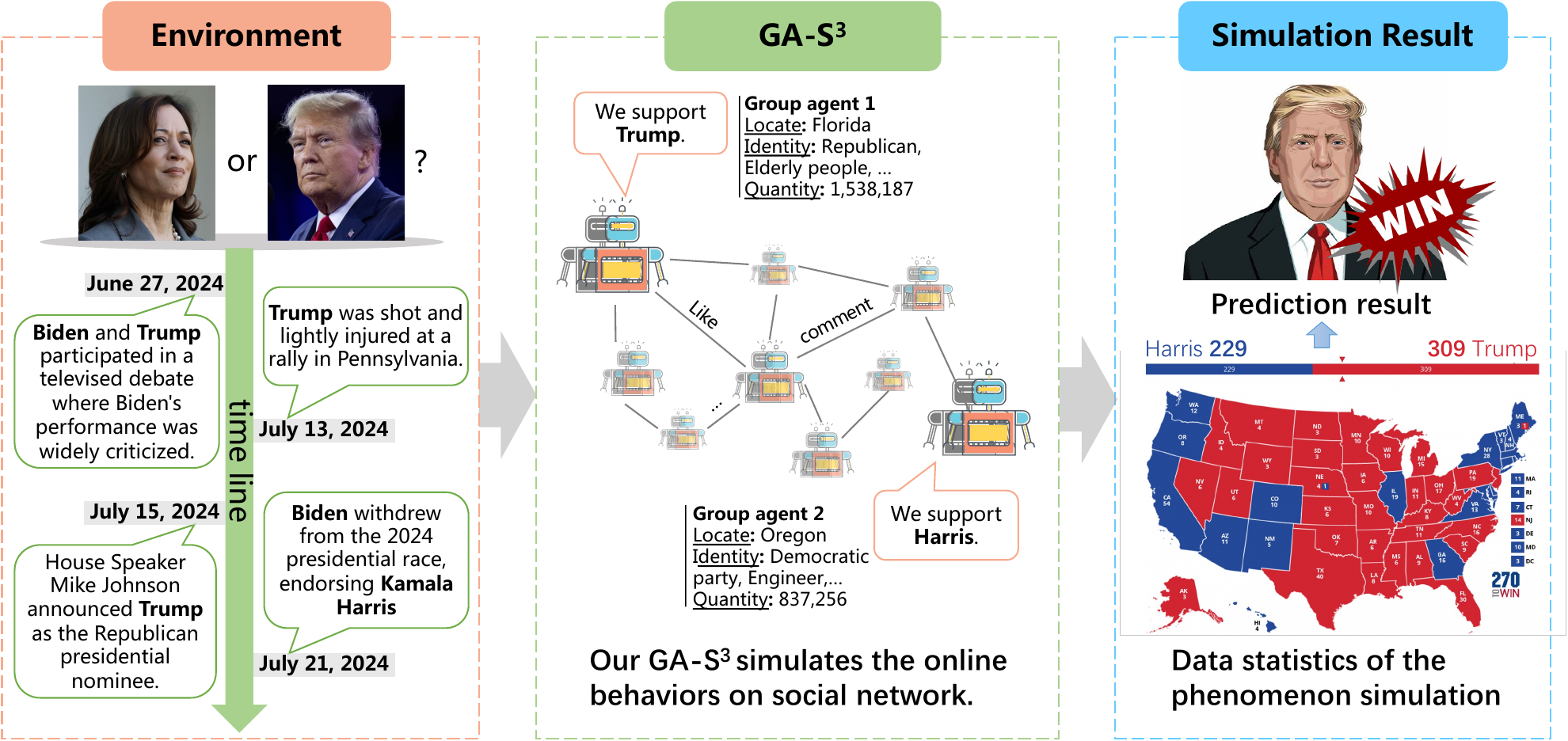}
\captionof{figure}{Prediction of the 2024 United States presidential election based on our social network simulation (GA-S\textsuperscript{3}). Our system predicted in July, 2024 that Trump would win with 309 electoral votes, which is remarkably close to the final result of 312 votes. 
}
\label{fig9:teaser}
\end{center}
    }]
}

\twocolumn[
{\Large\centerline{\textbf{GA-S\textsuperscript{3}: Comprehensive Social Network Simulation with Group Agents}}}
\vspace{1.0em}{\Large\centerline{\textbf{Appendix}}} 
\vspace{6.5em}
]

In this supplementary material, we describe our method in more detail in Section \ref{secA:Implement Detail}. We supplement our experiments with more details in Section B. Furthermore, in Section \ref{secC:US Elections}, we present a prediction for the 2024 United States elections in Figure~\ref{fig9:teaser}. Our GA-S\textsuperscript{3} system \textbf{predicted in July, 2024 that Trump would win with 309 electoral votes}, which is remarkably close to the final result of 312 votes.

\section{Implementation Details}
\label{secA:Implement Detail}

\begin{table*}[htbp]\footnotesize
\centering
\caption{Terminology Explanation}
\label{tab:rebuttal-3-terms}
\begin{tabular}{@{ }lp{2cm}p{8cm}p{4cm}@{}}
\toprule
\textbf{No.} & \textbf{Term} & \textbf{Explanation} & \textbf{Difference} \\
\midrule

1 & \textbf{Environmental Events} 
& Refers to the agent's perspective, focusing on how agents perceive and interpret events within their environment. 
& \multirow{2}*{\parbox[t]{4cm}{These terms describe events from different perspectives.}} \\

2 & \textbf{Online Events} 
& Refers to events that occur within the online social network, viewed from the dataset perspective. 
& \\
\midrule
3 & \textbf{Network Traffic Variations} 
& Refers to the fluctuations in the number of views an event receives over time, indicating its popularity and engagement across different periods. 
& - \\
\midrule
4 & \textbf{Population Weight} 
& Derived from real data, it influences the activity levels and interaction frequencies of groups.
& \multirow{4}*{\parbox[t]{4cm}{These terms describe fine-grain factors for real-world effects.}} \\
5 & \textbf{Group Characteristics} 
& Define the amplitude of emotional and attitudinal fluctuations within a group.
&  \\
6 & \textbf{\parbox[t]{1cm}{Emotion Fading}} 
& Refers to the gradual decline of emotions and attitudes over time. 
&  \\
7 & \textbf{Forgetting Probability} 
& Affects short-term memory, leading to the gradual fading of memories related to past perceptions and events. 
&  \\
\midrule
8 & \textbf{View} 
& Involves reading or viewing the event content.
& \multirow{4}*{\parbox[t]{4cm}{These terms describe the system's simulation of the real human action space.}} \\
9 & \textbf{Like} 
& Indicates interest or approval of the event by the user. 
&  \\
10 & \textbf{Comment} 
& Involves expressing thoughts or feedback on the event. 
&  \\
11 & \textbf{Share} 
& Refers to distributing or forwarding the event content to others. 
&  \\
12 & \textbf{Predict} 
& Involves projecting potential future outcomes based on the event's content and context. 
&  \\
\bottomrule
\end{tabular}
\end{table*}

\subsection{DTW Std Calculation Formula}
\label{DTW}
The standardization formula transforms time series data into a form with zero mean and unit variance to eliminate the effect of scale differences. The formula is:
\[
\tilde{X} = \frac{X - \mu(X)}{\sigma(X) + \epsilon}
\]
Where:
\begin{itemize}
    \item \(\mu(X)\) represents the mean of the time series \(X\),
    \item \(\sigma(X)\) represents the standard deviation of the time series \(X\),
    \item \(\epsilon\) is a small positive constant, typically \(10^{-8}\), used to avoid division by zero.
\end{itemize}

Dynamic Time Warping (DTW) calculates the minimum alignment distance between two time series. The formula is:
\[
\text{DTW}(\tilde{X}_1, \tilde{X}_2) = \sum_{n=1}^{N} d(\tilde{X}_{1,n}, \tilde{X}_{2,n})
\]
Where:
\begin{itemize}
    \item \(N\) is the number of time points,
    \item \(d(x, y)\) is the distance measure between two points, which can be:
    \begin{itemize}
        \item \text{Absolute error: } \(d(x, y) = |x - y|\),
        \item \text{Squared error: } \(d(x, y) = (x - y)^2\).
    \end{itemize}
\end{itemize}
In our study, the values for each day from the experiment are directly compared with the corresponding values from the ground truth.

Once the DTW distances are computed, the standard deviation (\(\text{std}\)) can be calculated as follows:
\[
\text{std} = \sqrt{\frac{1}{k} \sum_{i=1}^k (\text{DTW}_i - \mu_{\text{DTW}})^2}
\]
Where:
\begin{itemize}
    \item \(k\) is the number of DTW distances computed,
    \item \(\mu_{\text{DTW}}\) is the mean of the DTW distances.
\end{itemize}

\subsection{Terminology Explanation}
Here is a summary of the terms that appear in the article, as shown in Table \ref{tab:rebuttal-3-terms}.

\begin{table}[ht]\footnotesize
\centering
\caption{Agents and their population with characteristics in the education domain}
\label{tab6:agents_table}
\begin{tabular}{l c l}
\toprule
\textbf{Agents} & \textbf{Population} & \textbf{Characteristic} \\
\midrule
% Layer 1: Students
\textbf{\textit{Students}} & 58,030,769 & susceptible \\
% Layer 2: Students Categories
\textbf{Postgraduates} & 3,653,613 & calm \\
Doctor & 556,065 & calm \\
Master & 3,097,548 & ordinary \\
\textbf{Undergraduates} & 19,656,436 & susceptible \\
Bachelor & 19,656,436 & susceptible \\
\textbf{Vocation} & 34,720,720 & susceptible \\
Normal & 8,926,980 & susceptible \\
Short-cycle & 25,794,740 & susceptible \\
\midrule
% Layer 1: Teachers
\textbf{\textit{Teachers}} & 3,450,000 & calm \\
% Layer 2: Teachers Categories
\textbf{Educational Personnel} & 2,870,866 & calm\\
Full-time-Teachers & 2,005,188 & calm \\
Administrative-Personnel & 405,420 & calm \\
Supporting-Staff & 245,438 & ordinary \\
Workers & 122,982 & ordinary \\
Full-time-Researchers & 50,600 & calm \\
Other-Agency & 41,238 & ordinary \\
\textbf{Others Teachers} & 1,871,829 & calm \\
Part-time-Teachers & 453,302 & calm \\
Industry-Mentor & 405,037 & calm \\
Foreign-Teachers & 19,219 & calm \\
Retirees & 966,111 & calm \\
Affiliated-Teachers & 28,160 & ordinary \\
\bottomrule
\end{tabular}
\end{table}

\begin{figure*}[htb]
\centering
% 第一列：Views 的散点图
\subfloat[Scatter plot of views]{
    \includegraphics[width=0.253\linewidth]{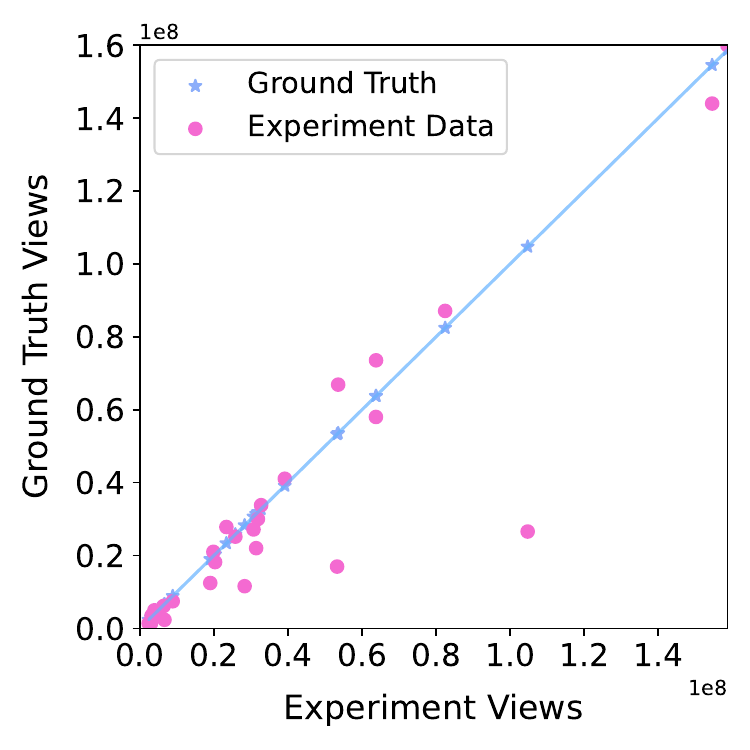}
}
\hfill
% 第二列：Views 的箱线图
\subfloat[Views box plot]{
    \includegraphics[width=0.207\linewidth]{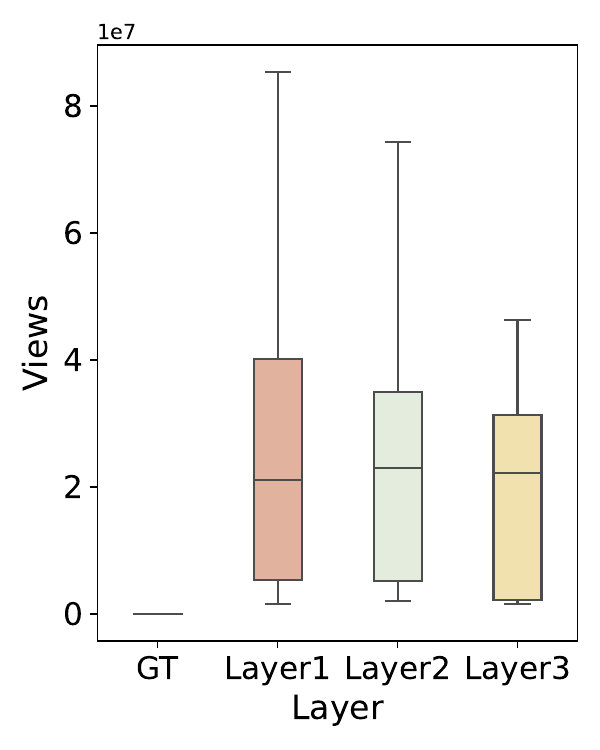}
}
\hfill
% 第三列：DTW distance 的散点图
\subfloat[DTW distance scatterplot]{
    \includegraphics[width=0.253\linewidth]{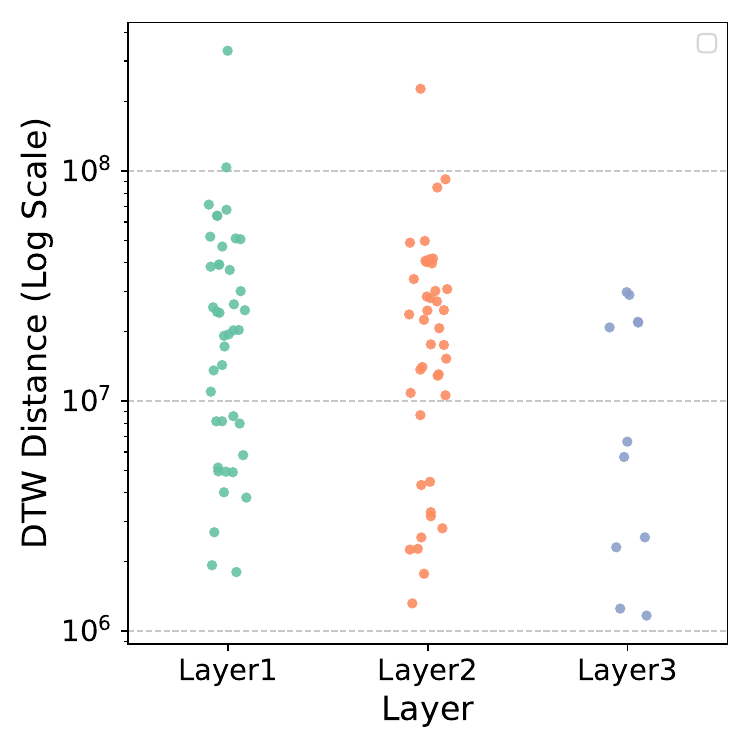}
}
\hfill
% 第四列：DTW distance 的箱线图
\subfloat[DTW distance boxplot]{
    \includegraphics[width=0.207\linewidth]{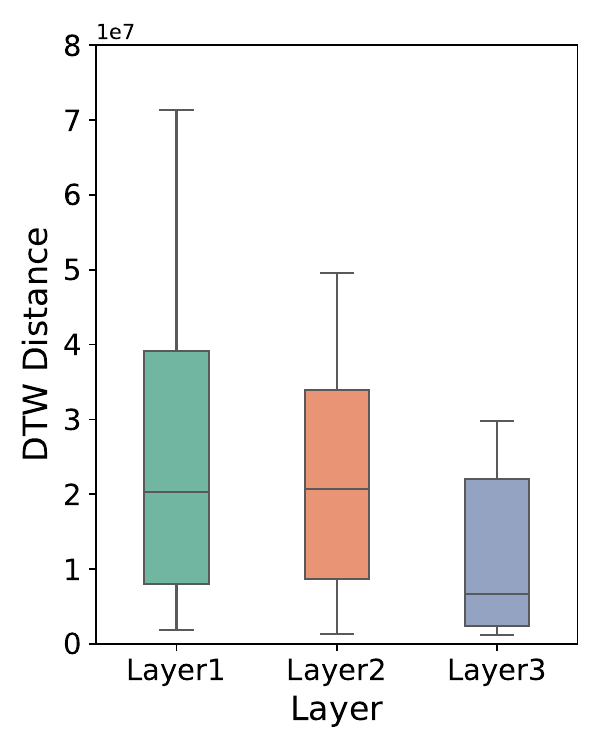}
}
\caption{
Comparison of experimental results across four visualizations. 
(a) Scatterplot of views compared to ground truth, illustrating their distribution. 
(b) Boxplot of views for different layers. 
(c) Scatterplot of DTW distances across layers, displayed on a logarithmic scale. 
(d) Boxplot of DTW distances by layer.
}
\label{fig3}
\end{figure*}

\subsection{Group Agents in Education Domain} 
Table \ref{tab6:agents_table} divides educational groups into three layers. The first layer includes students and teachers. The second layer categorizes students into postgraduates, undergraduates, and vocational students, and teachers into educational personnel and other teachers. The third layer further subdivides these groups based on specific roles, such as master’s students, full-time teachers, and industry mentors, each with distinct population sizes and emotional characteristics.

\subsection{Dataset Disclosure Statement}
Our \textbf{SNB} benchmark dataset captures diverse network traffic variations and a broad spectrum of news topics. While comprehensive, it can be further expanded for greater diversity. All data complies with the \textbf{General Data Protection Regulation (GDPR)}, ensuring user privacy through desensitization and anonymization of individuals in public social events.

\subsection{Key Prompts} 
Key prompts are presented in the following tables: Table \ref{tab7:group_find} and Table \ref{tab8:group_agent_generation} detail the hierarchical generation of group agents; Table \ref{tab9:agent_decision_instructions} illustrates the decision-reasoning processes of group agents; and Table \ref{tab10:agent_emotion_update} and Table \ref{tab11:agent_action_instructions} describe the emotional updates and predict actions of group agents, respectively.

\section{Additional Experiments}
\label{secB: Experiment Supplement}

\subsection{Visual Analysis of Metric}

Figure \ref{fig3} presents a comparison of experimental results using four visualizations. Figure \ref{fig3}(a) shows a scatter plot comparing the total Views of 30 events with real data, where the approximate linearity indicates a strong fit between our system and the actual data. Figure \ref{fig3}(b) provides a Views box plot, illustrating that a higher number of layers results in a more accurate fit. Figure \ref{fig3}(c) depicts a scatter plot of the DTW distance between the 7-day trend curves of the 30 events and the real results. As the number of layers increases, the DTW distance becomes smaller and more concentrated. This is further supported by the DTW box plot in Figure \ref{fig3}(d), which demonstrates that a greater number of layers reduces error.

\subsection{Demographic Influence on Model Generalization}

To investigate the influence of demographic data and macro social network structures on model generalization, additional experiments were conducted under two scenarios involving missing data: (1) data generated by a large language model (LLM), and (2) the absence of demographic data, as shown in Table \ref{tab:rabuttal-1-Demographic Generalization}.

\begin{table}[h!]\small
\centering
\caption{Performance comparison under different data conditions}
\label{tab:rabuttal-1-Demographic Generalization}
\begin{tabular}{@{ }lcccc@{}}
\toprule
\textbf{Scenario} & \textbf{MAPE} & \textbf{DTW (Std)} \\
\midrule
Real data & 17.17\% & 0.1925 \\
LLM-generated & 17.23\% & 0.1929 \\
Without data & 1421.71\% & 0.7921 \\
\bottomrule
\end{tabular}
\end{table}

The experimental results indicate that the model performs similarly when using LLM-generated data and real-world data. In contrast, the absence of demographic data significantly reduces the model’s accuracy. These findings suggest that the model exhibits strong generalization capabilities when demographic information is approximated through LLM-generated data.

\subsection{Inference Time Comparison with Existing Methods}

\begin{table}[h!]\small
\centering
\caption{Comparison of average processing time for different agent configurations}
\label{tab:rabuttal-2-time_comparison}
\begin{tabular}{@{ }lccc@{}}
\toprule
\textbf{Method} & \textbf{Number} & \textbf{Average Cost Time} \\
\midrule
ours-L1 & 2 agents & 90 seconds \\
ours-L2 & 4 agents & 200 seconds \\
ours-L3 & 16 agents & 13 minutes \\
$S^3$ & 1000 agents & 1 hour+ \\
\bottomrule
\end{tabular}
\end{table}

The average processing time per event in our method varies with the number of agents at each level. Compared to the $S^3$ method, our approach demonstrates significantly faster inference times across different agent configurations, as summarized in the table \ref{tab:rabuttal-2-time_comparison}.

\clearpage

% \subsection{Comparison with Other Large Language Models}

% We conducted additional experiments to compare the performance of our method with other large language models, including Qwen2-7b, as shown in Table \ref{tab:rabuttal-3-llm_comparison}. 

% \begin{table}[h!]\small
% \centering
% \caption{Performance comparison with large language models}
% \label{tab:rabuttal-3-llm_comparison}
% \begin{tabular}{@{ }lccc@{}}
% \toprule
% \textbf{Model} & \textbf{t-test} & \textbf{MAPE} & \textbf{DTW (Std)} \\
% \midrule
% LLaMA3 & 0.389 & 16.48\% & 0.1890 \\
% Qwen2  & 0.394 & 16.74\% & 0.1913 \\
% \bottomrule
% \end{tabular}
% \end{table}

\begin{figure*}[htb]
\centering
\subfloat[Biden VS Trump predictions]{\includegraphics[width=0.48\linewidth]{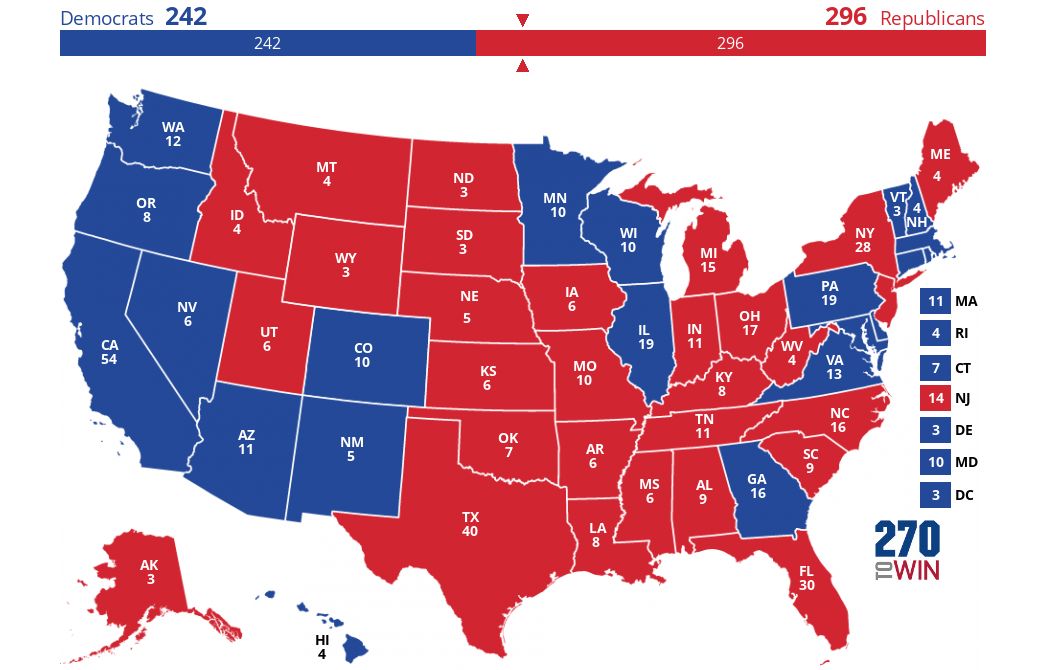}}
\hfill
\subfloat[Harris VS Trump predictions]{\includegraphics[width=0.48\linewidth]{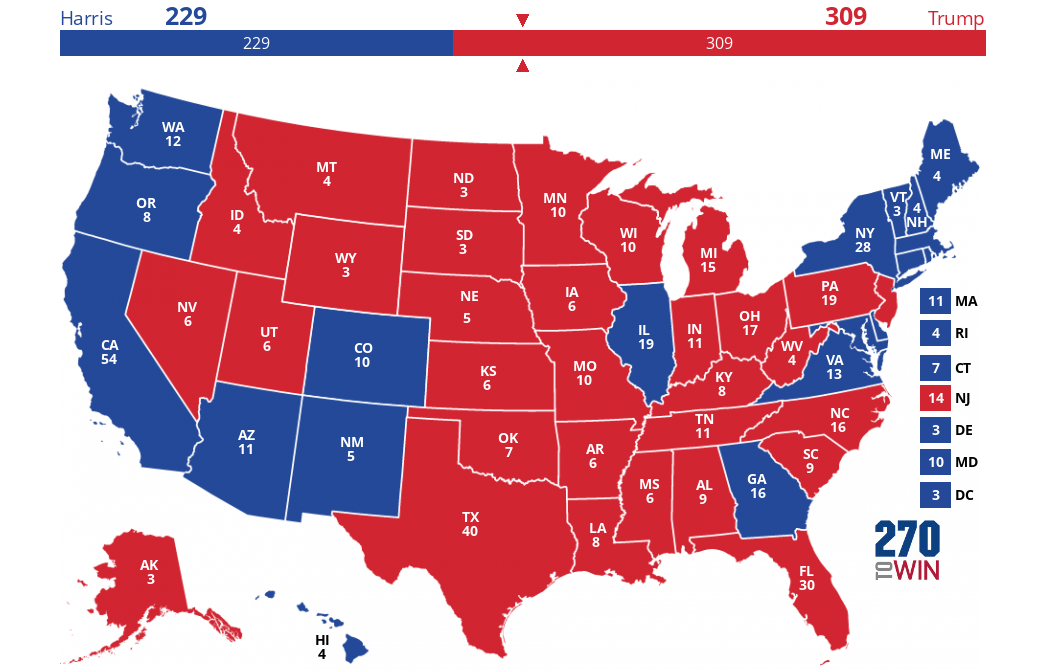}}
\caption{Statistics results of the simulation for the 2024 United States presidential election.}
\label{fig7}
\end{figure*}

\section{Predictions for 2024 US Elections}
\label{secC:US Elections}

\tcbset{
    colback=blue!5,    % 背景颜色（浅蓝）
    colframe=blue!75,  % 边框颜色（深蓝）
    fonttitle=\bfseries, % 标题字体样式
    title=Declaration,   % 框的标题
    width=0.48\textwidth,     % 宽度为页面宽度
    boxrule=0.5mm,        % 边框线条粗细
    arc=4mm,              % 圆角
}

\begin{tcolorbox}
\textbf{We declare that our system's prediction for the 2024 United States presidential election was made on July 25, 2024, and has not been adjusted or changed since then. Our submission to OpenReview on August 15, 2024, serves as evidence. This simulation is for academic research and discussion purposes only. The predictions and opinions included in this study are for reference only and do not represent the stance of the authors or the research team. These predictions should not be interpreted as definitive forecasts or guarantees.}
\end{tcolorbox}

\subsection{Prediction results}
The uncertainty surrounding \textbf{the 2024 United States presidential election} provides an opportunity to assess the predictive capabilities of our system.

On November 5, 2024, the United States will hold its 60th presidential election. This election features key figures such as Biden, whose performance \cite{Hadas_Gold_51_million_viewers} has been notable, Trump, who was involved in a shooting incident \cite{layne_2024_pop_trump}, and Harris, the Democratic nominee \cite{daniels_2024_nodate}. The goal of our system is to predict the election outcomes for state-level groups to identify the ultimate winner.

To address this series of news events, our virtual social network system selected citizens from all 51 states and created 51 group agents, each representing the population of a specific state. To facilitate predictions for each state's citizens, the system includes a prediction action. When the system receives new election activity, each state's group agent updates its emotions and attitudes and engages in interactions. After the interaction actions and the evolution of emotions and attitudes are completed, the system predicts election outcomes based on current emotional attitudes, interaction information, and the results from the 2020 election. The predictions, both before and after Biden's withdrawal, suggest that the Republican candidate Trump is likely to defeat the Democratic opponent, as shown in Figure \ref{fig7}.

\begin{table}[ht]\small
\centering
\begin{tabular}{lccc}
\toprule
\multicolumn{1}{c}{\textbf{Comparison}} & \multicolumn{1}{c}{\textbf{Candidate}} & \multicolumn{1}{c}{\textbf{Votes}} & \multicolumn{1}{c}{\textbf{Support Rate}} \\ \midrule
\multirow{2}{*}{Biden vs Trump} & Biden & 242 & 45.32\% \\ 
               & Trump & 296 & 54.68\% \\ \midrule
\multirow{2}{*}{Harris vs Trump} & Harris & 229 & 42.57\% \\ 
                & Trump & 309 & 57.43\% \\ 
\bottomrule
\end{tabular}
\caption{Comparison between the votes and support rates of presidential candidates}
\label{table4:comparison_candidates}
\end{table}

Before Biden announced his withdrawal, the system processed news from June 27 to July 15. The predicted results are illustrated in Figure \ref{fig7}(a), showing Trump leading Biden by 296 electoral votes. After Biden announced his withdrawal and endorsed Harris, the system generated a new prediction, as depicted in Figure \ref{fig7}(b). The updated prediction indicated that Trump would win in swing states such as Nevada (NV), Minnesota (MN), Wisconsin (WI), Michigan (MI), and New Jersey (NJ), securing 309 electoral votes. The predicted electoral votes and support rates are detailed in Table \ref{table4:comparison_candidates}. Our system forecasted that Trump's votes and support rate would be higher, with further increases expected following Biden's withdrawal and Harris's endorsement. This prediction aligns with the reactions of people in each state to the relevant news, demonstrating the scientific predictive capability of our group agent system.

\subsection{Scientific Control}
Before the announcement of the election results, numerous studies employed various methods to predict the outcome of the 2024 U.S. presidential election. 

\begin{figure*}[htb]
\centering
\subfloat[Actual 2024 election results]{\includegraphics[width=0.48\linewidth]{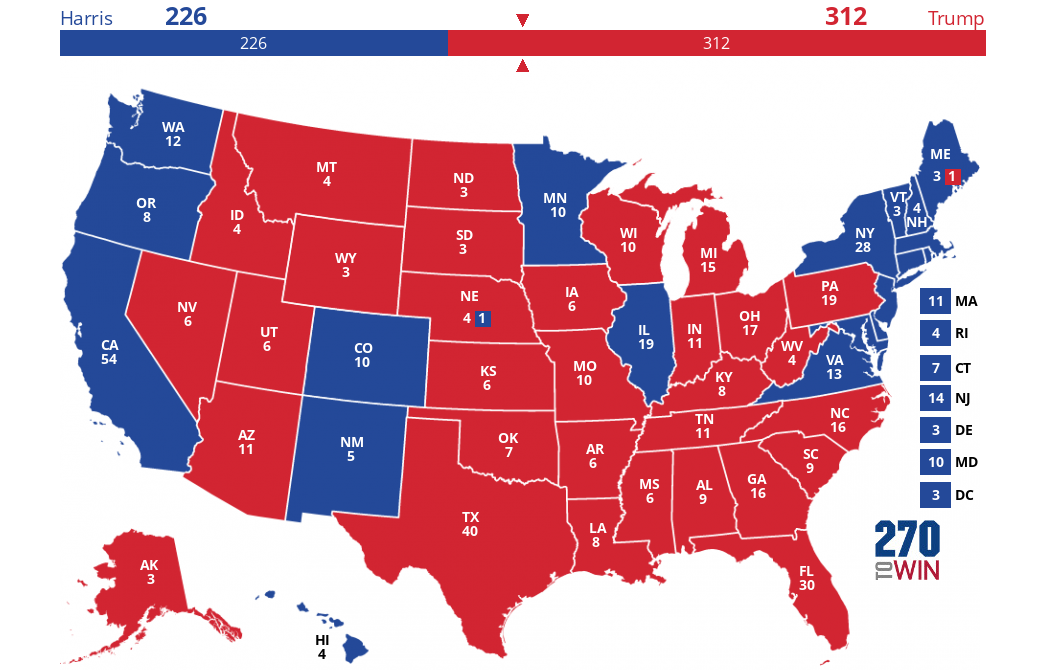}}
\hfill
\subfloat[Ours predictions]{\includegraphics[width=0.48\linewidth]{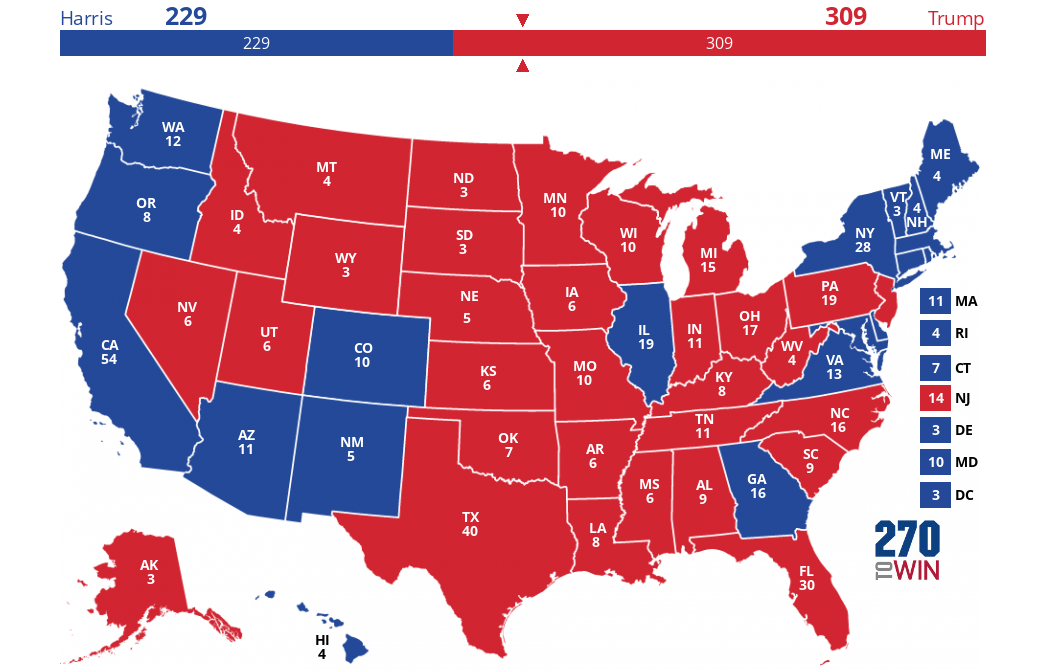}}
\hfill
\subfloat[ElectionSim predictions]{\includegraphics[width=0.48\linewidth]{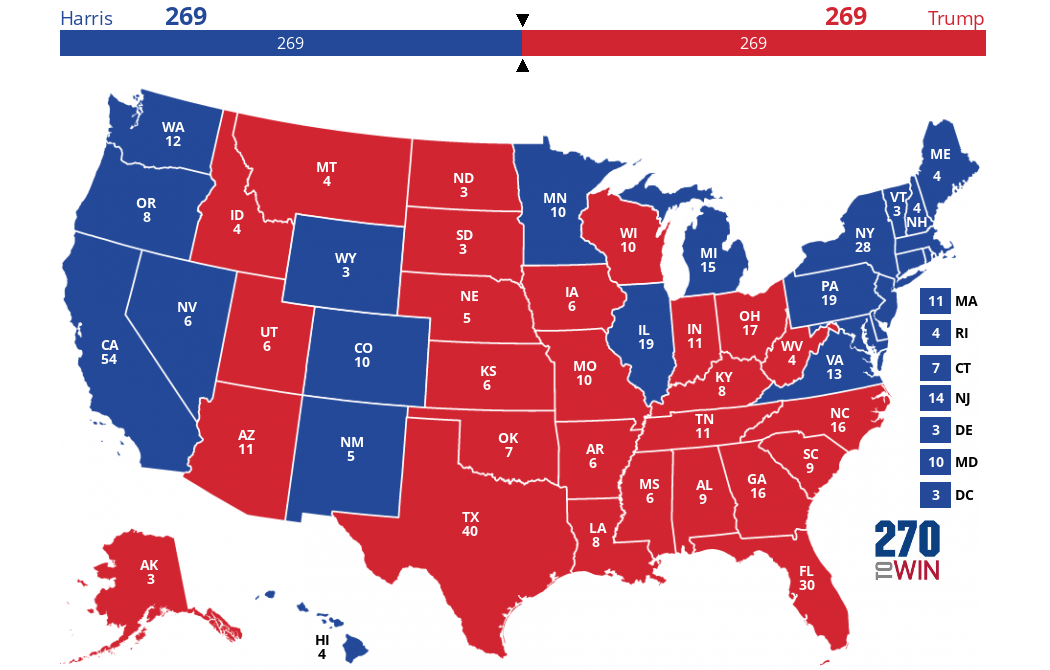}}
\hfill
\subfloat[Multi-step predictions]{\includegraphics[width=0.48\linewidth]{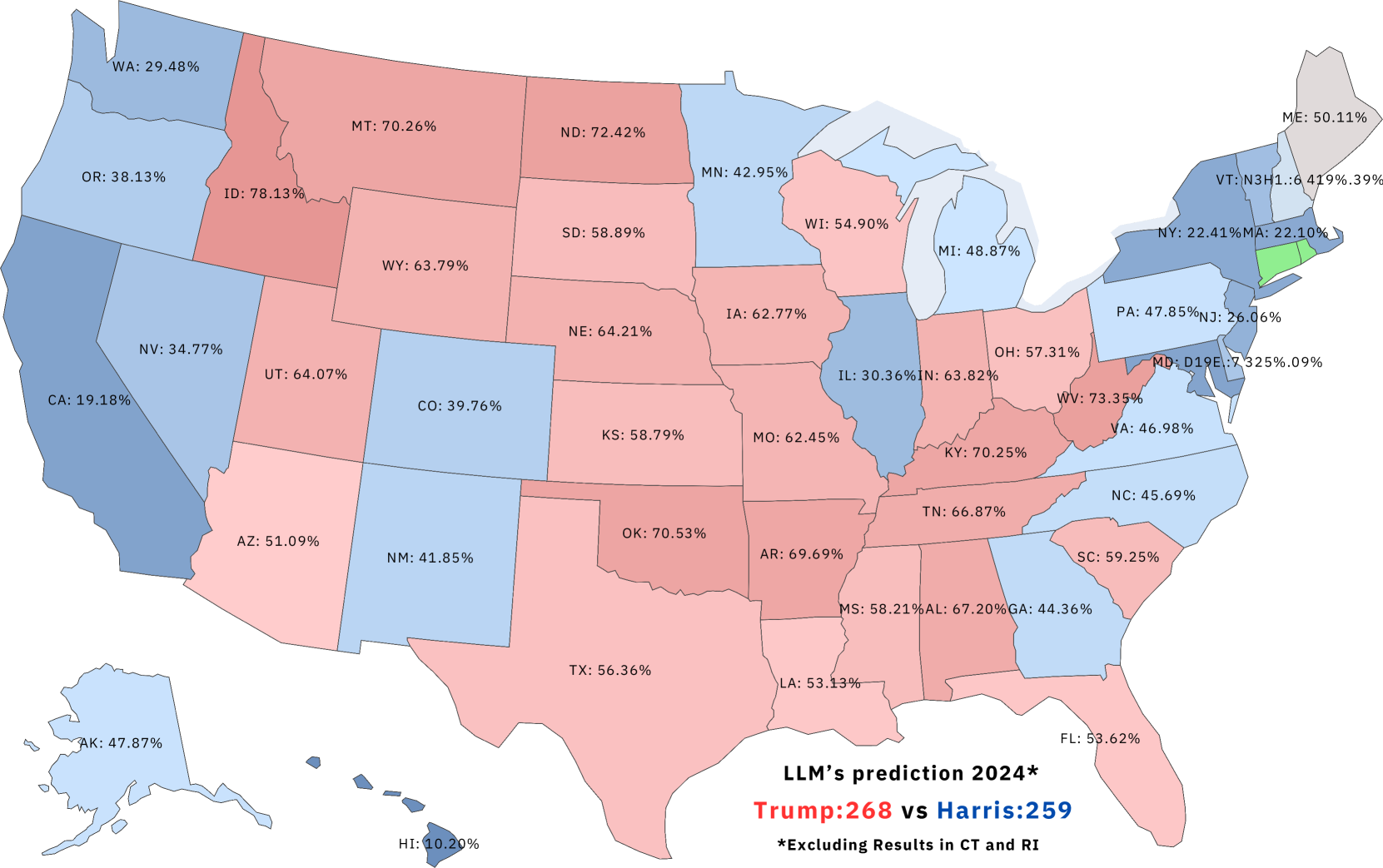}}
\caption{The actual results of the 2024 U.S. presidential election and the simulated statistical results using various methods.}
\label{fig8}
\end{figure*}

\textbf{ElectionSim} \cite{Electionsim-twitter-2024} is a framework based on large language models designed to simulate the 2024 U.S. presidential election, as illustrated in Figure \ref{fig8}(c). It accurately models voter preferences using a database containing millions of voters and incorporates data from the 2020 ANES survey and the 2022 U.S. Census to represent state-level population distributions. Evaluated using the presidential election benchmark (PPE), the simulation predicts Kamala Harris and the Democratic Party winning 8 out of 15 battleground states, while Donald Trump and the Republican Party are projected to secure 7, suggesting a slight advantage for the Democrats.

\textbf{Multi-step Reasoning with Large Language Models}\cite{Multi-step-2024} introduces a predictive framework that integrates synthetic data, real data, and a multi-step reasoning approach to enhance forecasting accuracy for the 2024 U.S. presidential election. The model predicts improved performance for Trump in swing states such as Wisconsin (WI), while projecting a narrow lead for Harris in key states like Pennsylvania (PA) and Michigan (MI). The forecast indicates a tight contest, with Trump potentially securing 268 electoral votes to Harris’s 259. However, incorporating data from additional states could tip the balance in Harris’s favor, giving her 270 votes.

\textbf{The actual 2024 United States elections}\cite{wiki-2024-elections-results} were held on Tuesday, November 5, 2024, as shown in Figure \ref{fig8}(a). In the presidential election, former Republican President Donald Trump, seeking a non-consecutive second term, defeated the incumbent Democratic Vice President Kamala Harris.

In comparison with other methods, ElectionSim predicts a slight advantage for Harris. While Multi-step forecasts Trump winning 268 electoral votes, it also suggests that the result is biased in favor of Harris, giving him 270 votes. Our system’s predicted outcome of 309 votes is closer to the actual result of 312 votes. The prediction results for swing states are compared in the Table \ref{table5:comparison_winner}. Both our method and ElectionSim achieve an accuracy of 86.7\% across 15 swing states, with three states predicted incorrectly. In contrast, the Multi-step approach demonstrates an accuracy of 66.7\%, mispredicting five swing states.

\begin{table*}[h!]
\centering
\caption{Simulation results for the 2024 presidential election in 15 battleground states. ($\star$ indicates correct prediction)}
\label{table5:comparison_winner}
\begin{tabular}{l c c c c}  % 定义表格列格式：两列左对齐，七列右对齐
\toprule  % 顶部粗线
State &  \multicolumn{4}{c}{Winner} \\  % 主标题行
\cmidrule(lr){2-5}   % Answering 和 User 部分的细线
& GT & Ours & ElectionSim & Multi-step \\  % 次级标题行
\midrule
Arizona       & Trump-Vance  & Harris-Walz  & Trump-Vance $\star$ & Trump-Vance $\star$ \\ 
Colorado      & Harris-Walz  & Harris-Walz $\star$ & Harris-Walz $\star$ & Harris-Walz $\star$ \\ 
Florida       & Trump-Vance  & Trump-Vance $\star$ & Trump-Vance $\star$ & Trump-Vance $\star$ \\ 
Georgia       & Trump-Vance  & Harris-Walz  & Trump-Vance $\star$ & Harris-Walz  \\ 
Iowa          & Trump-Vance  & Trump-Vance $\star$ & Trump-Vance $\star$ & Trump-Vance $\star$ \\ 
Michigan      & Trump-Vance  & Trump-Vance $\star$ & Harris-Walz  & Harris-Walz  \\ 
Minnesota     & Harris-Walz  & Trump-Vance  & Harris-Walz $\star$ & Harris-Walz $\star$ \\ 
Nevada        & Trump-Vance  & Trump-Vance $\star$ & Harris-Walz & Harris-Walz \\ 
New Hampshire & Harris-Walz  & Harris-Walz $\star$ & Harris-Walz $\star$ & Harris-Walz $\star$ \\ 
North Carolina & Trump-Vance  & Trump-Vance $\star$ & Trump-Vance $\star$ & Harris-Walz  \\ 
Ohio          & Trump-Vance  & Trump-Vance $\star$ & Trump-Vance $\star$ & Trump-Vance $\star$ \\ 
Pennsylvania  & Trump-Vance  & Trump-Vance $\star$ & Harris-Walz  & Harris-Walz  \\ 
Texas         & Trump-Vance  & Trump-Vance $\star$ & Trump-Vance $\star$ & Trump-Vance $\star$ \\ 
Virginia      & Harris-Walz  & Harris-Walz $\star$ & Harris-Walz $\star$ & Harris-Walz $\star$ \\ 
Wisconsin     & Trump-Vance  & Trump-Vance $\star$ & Trump-Vance $\star$ & Trump-Vance $\star$ \\ 
\midrule
Accuracy & - & 86.7\% & 86.7\% & 66.7\% \\
\bottomrule
\end{tabular}
\end{table*}

% 定义 tcolorbox 样式
\tcbset{
    colback=gray!10,    % 背景颜色（浅灰）
    colframe=black,     % 边框颜色（黑色）
    width=\textwidth,   % 宽度为页面宽度
    boxrule=0.5mm,      % 边框线条粗细
    arc=4mm,            % 圆角
    title=              % 移除标题
}

\begin{table*}[htbp]
\centering
\begin{tcolorbox}
\textbf{Instructions:} \\
You are an AI assistant specializing in generating hierarchical population group structures based on the provided country and domain. Use the given context to create a detailed tree-structured hierarchy that includes group names and corresponding numbers at each level.\\
Domain: \{domain\}      Country: \{country\} \\
\textbf{Your task is to generate a multi-level hierarchy for population groups, adjusting the structure based on the country and domain. Use the following format:} 
\begin{tabbing}
\hspace{1em} \= • \= First Layer (Domain-level Groups, denoted by \#\#): \\
\hspace{1em} \=    Broad categories representing the major population groups of the domain in the given field.\\
\hspace{1em} \= • \= Second Layer (Subgroups, denoted by 1. ** **): \\
\hspace{1em} \=    Specific subdivisions of each first-layer group.\\
\hspace{1em} \= • \= Third Layer (Detailed Breakdown, denoted by -): \\
\hspace{1em} \=    Granular breakdowns within each subgroup.
\end{tabbing}
\textbf{Example:} \\
For Country: CN (China) and Field: Education, a branch of the tree structure should be like this: 
\begin{tabbing}
\#\# Students: 58,030,769 \\
\hspace{1em} \= 1. **Postgraduates: 3,653,613** \\
\hspace{2em} \=  - Doctor: 556,065 \\
\hspace{2em} \=  - Master: 3,097,548 \\
\hspace{1em} \= 2. **Undergraduates: 19,656,436** \\
\hspace{2em} \=  - Bachelor: 19,656,436 \\
\hspace{1em} \= 3. **Vocational Undergraduate: 34,720,720** \\
\hspace{2em} \=  - Normal: 8,926,980 \\
\hspace{2em} \=  - Short-cycle: 25,794,740 
\end{tabbing}
\end{tcolorbox}
\caption{Find hierarchical group information prompt template via Kimi}
\label{tab7:group_find}
\end{table*}

\begin{table*}[htbp]
\centering
\begin{tcolorbox}
\textbf{Instructions:} \\
You are an AI assistant tasked with generating group agents and your process is as follows: \\
1. Identify and list all groups mentioned in the document. \\
2. Based on the identified groups and their associated templates, generate an agent for each group, ensuring no duplicates and that all groups are generative. \\
Answer Format: 
\begin{tabbing}
\hspace{1em} \= agent \= \{n\}: \= (nth agent) \\
\hspace{1em} \= id: \= \{group\}-agents \\
\hspace{1em} \= description:  \= \parbox[t]{0.7\textwidth}{Representing \{number\} \{country\} \{group\}, reflecting their emotions, attitudes, and possible actions in response to the news.} \\
\hspace{1em} \= characteristic:  \= \{susceptible/ordinary/calm\} population
\end{tabbing}
3. Follow the template below strictly, filling in the \{group\}, \{number\}, and \{country\} fields based on the contextual input. \\

\end{tcolorbox}
\caption{Group agents generation template}
\label{tab8:group_agent_generation}
\end{table*}

\begin{table*}[htbp]
\centering
\begin{tcolorbox}[title=] % 空标题以避免显示任何标题
\textbf{System:} \\
You are \{agent\_name\}, \{agent\_description\}. \\
You are in the social network world: \{world\_description\}.
\begin{tabbing}
perception: \\
\hspace{1em} \= • \= Time: \{day\_n\} \\
\hspace{1em} \= • \= Event State: \{event\_state\} \\
% \hspace{1em} \= • \= Other Group Agents: \{others\_state\} \\
Your State: \\
\hspace{1em} \= • \= Previous Memory: \{memory\}  \\
\hspace{1em} \= • \= Previous State: \{previous\_state\} \\
\hspace{1em} \= • \= Current Emotion: \{emotions\} \\
\hspace{1em} \= • \= Current Attitude: \{attitudes\} 
\end{tabbing}
Action Options: \\
You can choose from the following available actions: \{available\_actions\} 

\textbf{Instructions:} \\
 1. Use decision-making reasoning to choose your actions based on factors such as perception and your status. This action must be one of the available actions based on the previous context. Also, explain why. \\               
 2. Answers must follow the following format: 
 \begin{tabbing}
\hspace{1em} \= Action: \{Action name\} \\
\hspace{1em} \= Reason: \{reason\} \\
\hspace{1em} \= Updated plan: \{List available actions with serial numbers\}
 \end{tabbing}
\end{tcolorbox}
\caption{Group agents decision-making prompt template}
\label{tab9:agent_decision_instructions}
\end{table*}

\begin{table*}[htbp]
\centering
\begin{tcolorbox}[title=] % 空标题以避免显示任何标题
\textbf{System:} \\
You are \{agent\_name\}, \{agent\_description\}. \\
You are in the social network world: \{world\_description\}.
\begin{tabbing}
perception: \\
\hspace{1em} \= • \= Time: \{day\_n\} \\
\hspace{1em} \= • \= Event State: \{event\_state\} \\
% \hspace{1em} \= • \= Other Group Agents: \{others\_state\} \\
Your State: \\
\hspace{1em} \= • \= Previous Memory: \{memory\}  \\
\hspace{1em} \= • \= Previous State: \{previous\_state\} \\
\hspace{1em} \= • \= Emotion Fading: \{emotion\_fading\} 
\end{tabbing}
\textbf{Instructions:} \\
1.	Update your emotions and attitudes: Update your emotions and attitudes based on your perception and status, taking into account the current time and emotion fading.  \\
2.	Event cycle pattern: In a typical event cycle, emotions will initially surge, then quickly decline, and eventually stabilize. Some explosive events may have a second emotional peak. Attitudes tend to follow a similar pattern. \\
3.	Response Template: 
\begin{tabbing}
\hspace{1em} \= emotions: \{ 'happiness': (), 'sadness': (), 'anger': () \} \\
\hspace{1em} \= attitudes: \{ 'optimism': (), 'pessimism': () \} 

\end{tabbing}
\end{tcolorbox}
\caption{Group agents update emotions and attitudes prompt template}
\label{tab10:agent_emotion_update}
\end{table*}

\begin{table*}[htbp]
\centering
\begin{tcolorbox}[title=] % 空标题以避免显示任何标题
\textbf{System:} \\
You are \{agent\_name\}, \{agent\_description\}. \\
You are in the social network world: \{world\_description\}.
\begin{tabbing}
perception: \\
\hspace{1em} \= • \= Time: \{day\_n\} \\
\hspace{1em} \= • \= Event State: \{event\_state\} \\
% \hspace{1em} \= • \= Other Group Agents: \{others\_state\} \\
Your State: \\
\hspace{1em} \= • \= Previous Memory: \{memory\}  \\
\hspace{1em} \= • \= Previous State: \{previous\_state\} \\
\hspace{1em} \= • \= Forgetting Probability: \{forgetting\_probability\} \\
\hspace{1em} \= • \= Current Emotion: \{emotions\} \\
\hspace{1em} \= • \= Current Attitude: \{attitudes\} 
\end{tabbing}
\textbf{Instructions:} \\
\textbf{Task: Predict daily engagement metrics} 
\begin{tabbing}
1. Daily reading forecast: \\
\hspace{1em} \parbox[t]{0.95\textwidth}{• Based on your perception and status, consider the popularity of the event, the current date, and the forgetting probability, and estimate how many people in your group have viewed the event.} \\[1em]
2. General engagement pattern: \\
\hspace{1em} • Views: \\
\hspace{2em} • Must be at least one order of magnitude higher than likes. \\
\hspace{2em} \parbox[t]{0.95\textwidth}{• Due to the forgetfulness effect, views gradually diminish over time, and explosive events may have a second peak of views, but less than the first peak of views.} \\
\hspace{1em} • Likes, comments, and shares: \\
\hspace{2em} • Likes usually exceed comments and shares. \\
\hspace{2em} • For news that sparks heated discussions, comments or shares may exceed likes. \\[1em]
3. Forecast format: \\
\hspace{1em} \= \textbf{Date:} YYYY-MM-DD \\
\hspace{1em} \= \textbf{Views:} \{predicted\_views\} \\
\hspace{1em} \= \textbf{Likes:} \{predicted\_likes\} \\
\hspace{1em} \= \textbf{Comments:} \{predicted\_comments\} \\
\hspace{1em} \= \textbf{Shares:} \{predicted\_shares\}
\end{tabbing}
\end{tcolorbox}
\caption{Group agents predict daily engagement metrics action prompt template}
\label{tab11:agent_action_instructions}
\end{table*}

\end{document}